\documentclass[aps,prb,twocolumn,superscriptaddress,nolongbibliography]{revtex4-2}
\usepackage{graphicx}
\usepackage{notes2bib}
\usepackage{amssymb,amsmath}
\usepackage{siunitx}
\usepackage{comment}
\usepackage{xcolor}
\usepackage[utf8]{inputenc}

\begin{document}


\title{Orbital-Induced Crossover of the Fulde-Ferrell-Larkin-Ovchinnikov Phase into Abrikosov-like States}


\author{Tommy Kotte} 
\email[]{t.kotte@hzdr.de}
\affiliation{Hochfeld-Magnetlabor Dresden (HLD-EMFL) and W\"urzburg-Dresden Cluster of Excellence ct.qmat, Helmholtz-Zentrum Dresden-Rossendorf, 01328 Dresden, Germany}
\author{Hannes Kühne}
\affiliation{Hochfeld-Magnetlabor Dresden (HLD-EMFL) and W\"urzburg-Dresden Cluster of Excellence ct.qmat, Helmholtz-Zentrum Dresden-Rossendorf, 01328 Dresden, Germany}
\author{John A Schlueter}
\affiliation{Materials Science Division, Argonne National Laboratory, Argonne, IL 60439, USA}
\affiliation{Division of Materials Research, National Science Foundation, Alexandria, VA 22314, USA}
\author{Gertrud Zwicknagl}
\affiliation{Institute for Mathematical Physics, Technische Universität Braunschweig, 38106 Braunschweig, Germany}
\author{J. Wosnitza}
\affiliation{Hochfeld-Magnetlabor Dresden (HLD-EMFL) and W\"urzburg-Dresden Cluster of Excellence ct.qmat, Helmholtz-Zentrum Dresden-Rossendorf, 01328 Dresden, Germany}
\affiliation{Institut f\"ur Festk\"orper- und Materialphysik, TU Dresden, 01062 Dresden, Germany}

\date{\today}

\begin{abstract}
The Fulde-Ferrell-Larkin-Ovchinnikov (FFLO) state can emerge in superconductors for which the orbital critical field exceeds the Pauli limit. Here, we present angular-resolved specific-heat data of the quasi-two-dimensional organic superconductor $\kappa$-(ET)$_2$Cu(NCS)$_2$, with a focus on high fields in the regime of the FFLO transition. For an increasing out-of-plane tilt of the applied magnetic field, which leads to an increase of orbital contributions, we found that the nature of the superconducting transition changes from second to first order and that a further transition appears within the high-field superconducting phase. However, the superconducting state above the Pauli limit is stable for field tilt of several degrees. Since any finite perpendicular component of the magnetic field necessarily leads to quantization of the orbital motion, the resulting vortex lattice states compete with the modulated order parameter of the FFLO state leading to complex high-field superconducting phases. By solving the linearized self-consistency equation within weak-coupling BCS theory, we show that our results are clear experimental evidence of an orbital-induced transformation of the FFLO order-parameter into Abrikosov-like states of higher Landau levels.
\end{abstract}


\maketitle


In spin-singlet type-II superconductors, the superconducting (SC) state is suppressed if an applied magnetic field reaches one of two limits: (i) the orbital critical field, $H_{orb}$, at which the Cooper pairs are broken by the Lorentz force \cite{tinkham2004}, or (ii) the Pauli-limiting field, $H_{P}$, at which the binding energy of the Cooper pairs is compensated by Zeeman splitting \cite{Chandrasekhar1962,Clogston1962}. In most materials, $H_{P}$ exceeds $H_{orb}$ considerably, and the normal-conducting (NC) state is restored at $H_{orb}$ via a second-order phase transition, which is well described by the Werthamer-Helfand-Hohenberg (WHH) model \cite{Werthamer1966}. 

However, $H_{orb}$ can be significantly enhanced to above the Pauli limit for superconductors yielding quasiparticles of high effective masses or for those with a pronounced anisotropy \cite{Matsuda2007,Zwicknagl2010}. Fulde and Ferrell \cite{Fulde1964}, as well as Larkin and Ovchinnikov \cite{Larkin1964}, have independently shown that, in this case, the suppression of superconductivity at $H_{P}$ may be avoided by the formation of Cooper pairs with a finite center-of-mass momentum and an associated spatial modulation of the SC order parameter. Although an anisotropic Fermi surface can stabilize this Fulde-Ferrell-Larkin-Ovchinnikov (FFLO) state \cite{Shimahara1997}, its formation requires the material to be in the clean limit \cite{Takada1970} and a Maki parameter $\sqrt{2}H_{orb}/H_{P}$ larger than \SI{1.8}{} \cite{Gruenberg1966}. 

Quasi-two-dimensional (Q2D) organic superconductors \cite{Brown2015,Wosnitza2007} are among the few known materials that fulfill these conditions \cite{Matsuda2007,Beyer2013a,Wosnitza2018}. Consequentially, the FFLO state was first experimentally verified, by thermodynamic means,  in this class of materials \cite{Lortz2007,Beyer2012,Agosta2017}. Thorough investigations of the high-field properties established some hallmarks of the FFLO phase, i.e., a reincrease of the SC phase boundary at low temperatures \cite{Agosta2012,Lortz2007,Bergk2011,Tsuchiya2015,Beyer2012,Sugiura2019}, the appearance of a first-order transition close to $H_{P}$ \cite{Sugiura2019,Agosta2017}, and an enhancement of the $T_\mathrm{1}$-relaxation rate, measured by nuclear magnetic resonance, in the vicinity of the FFLO phase boundary \cite{Mayaffre2014,Koutroulakis2016}. 

The continuing search for these experimental signatures led to recent reports of FFLO physics in multiple classes of superconductors, including heavy-fermion compounds \cite{Lin2020,Kitagawa2018}, iron-based superconductors \cite{Cho2017,Kasahara2020,Kasahara2021}, and a transition-metal dichalcogenide \cite{Cho2021}. Still, clear experimental confirmations of the FFLO phase remain rare. A major difficulty lies in the fact that the properties of the FFLO state are non-universal depending upon the normal-state quasiparticles and their interactions \cite{Burkhardt1994}. In addition, the modulation of the SC order parameter can be masked by competing effects arising in high fields, such as by spin density waves in the $Q$ phase of CeCoIn$_5$ \cite{Lin2020}. Finally, the inevitable presence of orbital effects leads to deviations of the order parameter from the original FFLO prediction, which assumes a one-dimensional modulation \cite{Bulaevskii1973,Shimahara1997,Buzdin1996,Houzet2000}. The resulting effects on the experimental signatures of the FFLO state are not well understood at this time. Q2D organic superconductors offer the unique possibility to study the evolution of the FFLO phase with increasing orbital contribution, since $H_{orb}$ can effectively be controlled by tilting the applied field away from the SC planes \cite{Bulaevskii1973,Shimahara1997}. 

In this Letter, we present angular-resolved specific-heat data of the Q2D organic superconductor \mbox{$\kappa$-(ET)$_2$Cu(NCS)$_2$} [with ET = bis(ethylenedithio)-tetrathiafulvalene], focusing on the orbital suppression of the high-field SC phase that is well established as a realization of the FFLO state \cite{Lortz2007,Agosta2017,Wright2011,Mayaffre2014,Bergk2011,Agosta2012,Tsuchiya2015,Fortune2018}. \textcolor{black}{We found the NC-FFLO transition to be of second order for fields applied precisely in plane, where orbital effects are negligible \cite{Lebed2018}. Rotating the field away from the in-plane orientation leads to an increase of orbital effects and changes the nature of the transition to first order.} Before the high-field (FFLO) superconductivity is finally suppressed by orbital effects, our data indicate a transition to a further SC state. All observations are in line with theoretical predictions of a successive conversion of the FFLO order parameter into an Abrikosov-like one of higher-order Landau levels by increasing orbital contributions \cite{Bulaevskii1973,Shimahara1997}, and represent a general phenomenology of Q2D Pauli-limited superconductors. 

We measured the specific heat of a \mbox{$\kappa$-(ET)$_2$Cu(NCS)$_2$} single crystal ($3\times 1 \times\SI{0.6}{\milli\meter^3}, \SI{4.42}{\milli\gram}$), grown by the standard electrolytic method \cite{Urayama1988}. The sample was attached to a \SI{360}{\degree} piezo-driven rotator and placed into a \SI{22}{\tesla} cryomagnet with $^3$He insert \cite{NoteSuppl}. The rotation axis was approximately parallel to the crystallographic $b$ axis. We determined the specific heat, $C$, using a continuous relaxation method that allows to detect thermal hystereses and is particularly sensitive to first-order transitions \cite{Wang2001,Beyer2012,NoteSuppl}. We aligned the field in plane by maximizing the transition temperature $T_{c}$ as function of the rotation angle $\alpha$ of the sample in a field of \SI{14}{T}. 
\begin{figure}[tb]
	\includegraphics{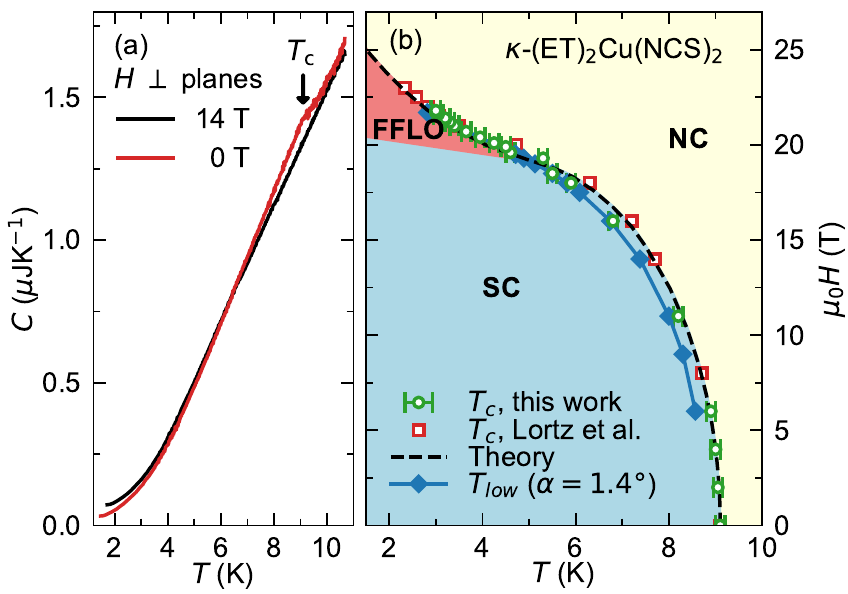}
	\caption{\label{Fig_1} (a) Heat capacity of the $\kappa$-(ET)$_2$Cu(NCS)$_2$ sample in zero field and at \SI{14}{\tesla} applied perpendicular to the ET-layers. The latter represents the NC heat capacity. (b) SC phase diagram of $\kappa$-(ET)$_2$Cu(NCS)$_2$ for in-plane fields. The theory curve (dashed line) is calculated under the assumption of an isotropic in-plane Fermi velocity. Data taken from Ref. \cite{Lortz2007} are shown for comparison. $T_{low}$ corresponds to the position of the first-order peak observed for $\alpha=\SI{1.4}{\degree}$ in $\Delta C$ [Fig. \ref{Fig_2}(b)].}
\end{figure}

Figure \ref{Fig_1}(a) shows the heat capacity of the sample in zero field, as well as in the NC state for a field of $\SI{14}{T}$, applied perpendicular to the ET layers. \textcolor{black}{The SC transition at $T_c\approx\SI{9.1}{\kelvin}$ appears as a broad shoulder in the zero-field data. The data agree with literature \cite{Andraka1989,Mueller2002,Wosnitza2003,Lortz2007}}.
\begin{figure}[tb]
	\includegraphics{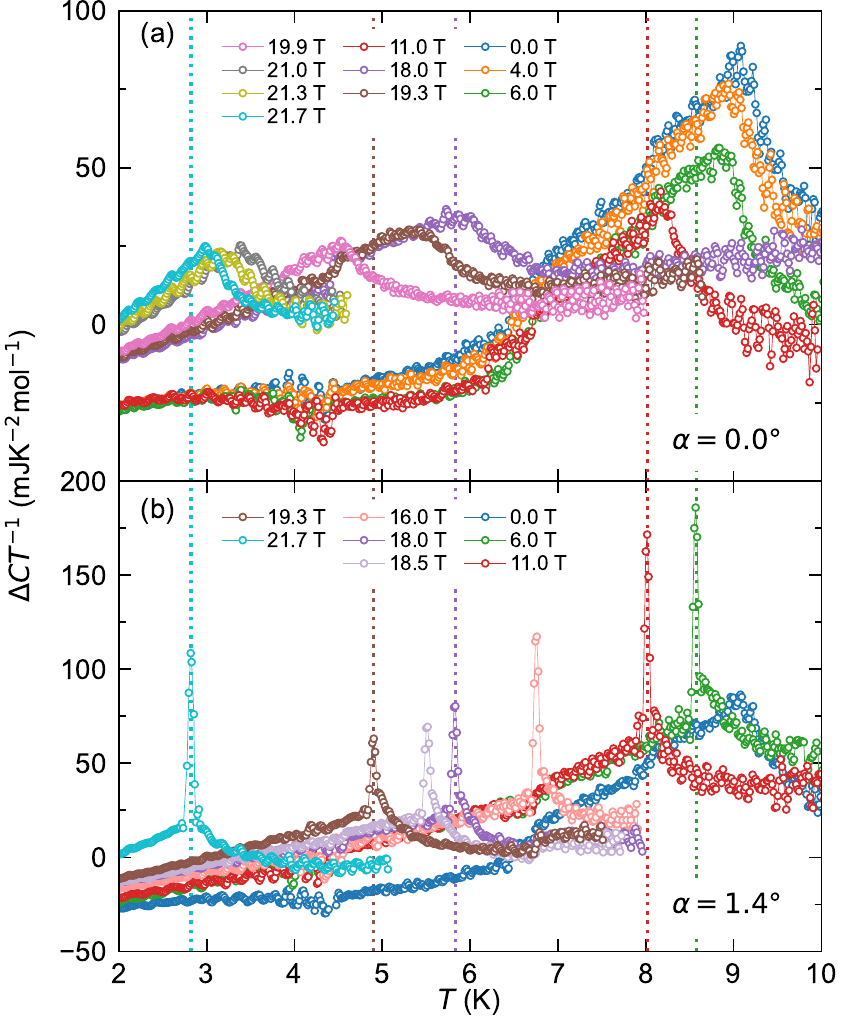}
	\caption{\label{Fig_2}Temperature dependence of $\Delta C/T$ of $\kappa$-(ET)$_2$Cu(NCS)$_2$ for different fields applied (a) parallel to the ET layers and (b) with an off-alignment of $\alpha=\SI{1.4}{\degree}$. $T_{low}$ is marked by vertical dashed lines for comparison.}
\end{figure}

Figure \ref{Fig_2}(a) shows the specific heat of $\kappa$-(ET)$_2$Cu(NCS)$_2$ divided by $T$ for different fields between \SI{0}{} and \SI{21.7}{\tesla}, applied parallel to the ET layers. The NC specific heat, shown in Fig. \ref{Fig_1}(a), is subtracted from the data, leaving only the change in the electronic specific heat $\Delta C$ stemming from the SC transition. For all fields, the SC transition manifests as a $\lambda$-like anomaly in $\Delta C/T$, and shifts to lower temperatures with increasing field. 

Figure \ref{Fig_1}(b) shows the phase diagram extracted from the peak positions in $\Delta C$. Up to $\SI{6}{\tesla}$, $T_{c}$ shifts only slightly with field. Applying the WHH extrapolation \cite{Werthamer1966}, we estimate the orbital critical field from the initial slope $dH_\mathrm{c}/dT|_{T=T\mathrm{c}}$ as about $\SI{260}{\tesla}$, exceeding $H_P$  \cite{Lortz2007} by more than an order of magnitude. \textcolor{black}{At intermediate fields, the slope of the phase boundary decreases up to \SI{19.3}{\tesla}. However,} above about $\SI{20}{\tesla}$, an upturn of the phase boundary below \SI{4}{K} occurs, representing a hallmark of the FFLO state \cite{Shimahara1997,Vorontsov2005}. This phase diagram is in agreement with specific-heat measurements by Lortz \textit{et al.}, although they observed additional peaks in the specific heat above \SI{21}{\tesla} \cite{Lortz2007}. 

\begin{figure*}[t]
	\includegraphics{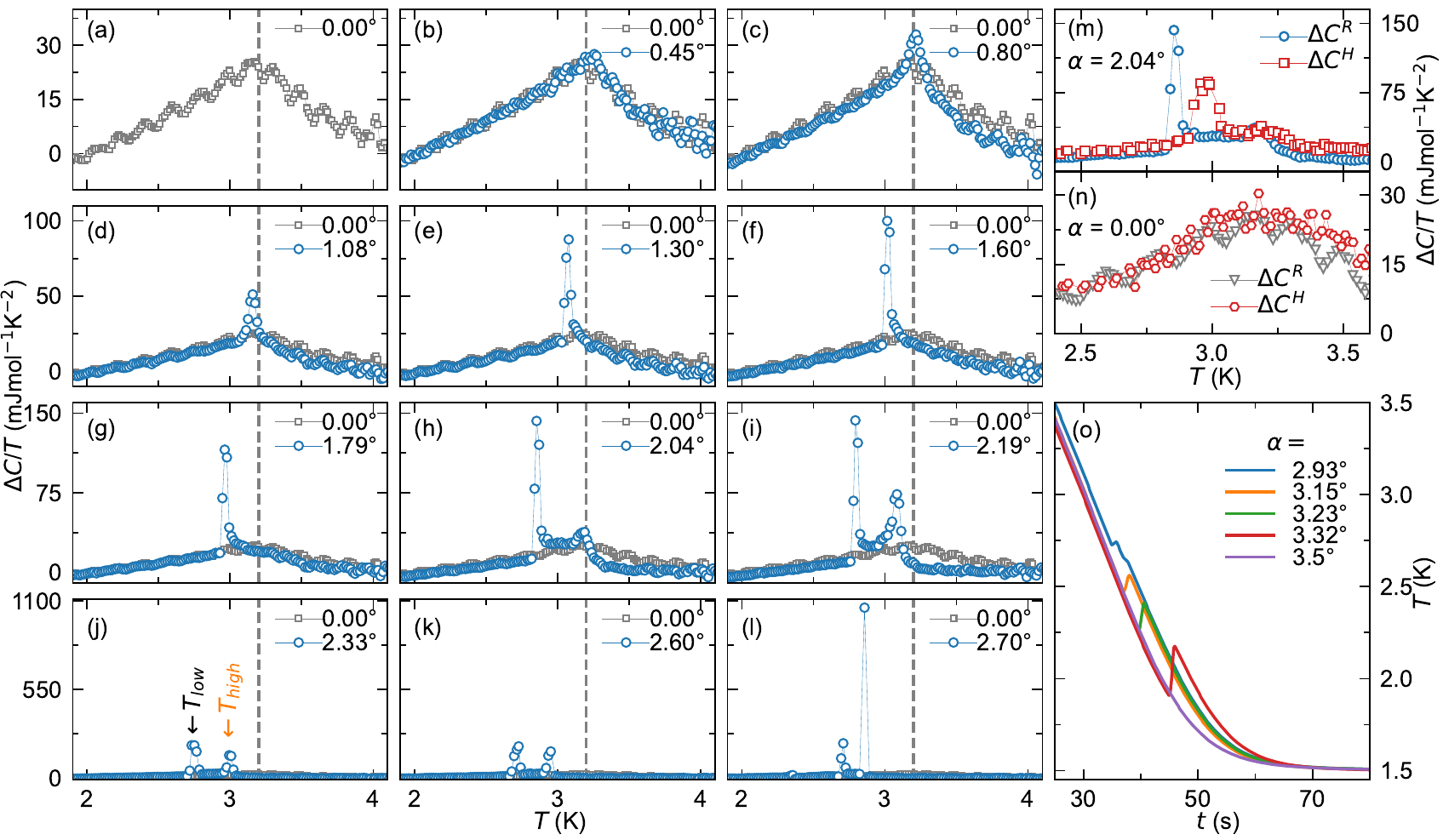}
	\caption{\label{Fig_4}(a)-(l) $\Delta C/T$ of $\kappa$-(ET)$_2$Cu(NCS)$_2$ at \SI{21.3}{\tesla} for selected angles $\alpha$. The vertical gray dashed line marks $T_{c}$ at \SI{21.3}{\tesla} and $\alpha=\SI{0}{\degree}$. Note the different scalings in each row. $\Delta C$, measured during heating (superscript H) and relaxation (superscript R), shows pronounced hysteresis for off-alignment (m), but not for in-plane field (n). (o) Thermal-relaxation curves for $\alpha$ between \SI{2.9}{\degree} and \SI{3.5}{\degree} revealing pronounced supercooling effects.}
\end{figure*}

Turning to the question of the stability of the FFLO state against orbital effects, we performed angular-resolved specific-heat measurements at \SI{21.3}{\tesla} (Fig. \ref{Fig_4}). In panels (a)-(l) of Fig. \ref{Fig_4}, $\Delta C/T$ is plotted for increasing $\alpha$, which is defined as the angle between $H$ and the ET layers. For an in-plane field orientation [Fig. \ref{Fig_4}(a)], a broad (triangular) peak marks the transition from the NC to the FFLO state at $T_{c}=\SI{3.2}{\kelvin}$ [see also Fig. \ref{Fig_2}(a)]. For slight off-orientations up to $\alpha\approx\SI{0.5}{\degree}$, the shape of the specific-heat anomaly and $T_c$ do not change [Fig. \ref{Fig_4}(b)]. For $\alpha$ between \SI{0.8}{\degree} and \SI{1.8}{\degree}, the broad transition is superimposed by a peak at $T_{low}$ [Figs. \ref{Fig_4}(c)-\ref{Fig_4}(g)]. This feature, first evolving at $T_{low}\approx T_{c}$ for $\alpha=\SI{0.8}{\degree}$, sharpens and shifts to lower temperature with increasing $\alpha$. Above about \SI{2.0}{\degree}, a second peak appears at $T_{high}$ [Fig. \ref{Fig_4}(h)-\ref{Fig_4}(l)]. For $\alpha=\SI{2.04}{\degree}$, $T_{high}$ is exactly at the in-plane $T_c$. With further increasing $\alpha$, both peaks sharpen considerably and shift down in temperature. Two observations are remarkable: (i) As shown exemplarily for $\alpha=\SI{2.04}{\degree}$ in Fig. \ref{Fig_4}(m), both peaks are hysteretic when comparing the heating (superscript $H$) and the relaxation ($R$) branches of the continuous specific-heat measurements. In contrast, for in-plane alignment [Fig. \ref{Fig_4}(n)], no hysteresis is observed. (ii) The low-temperature peak superimposes the broad triangular peak that signals the SC transition and stretches to at least \SI{4}{\kelvin} for $\alpha\le\SI{1.8}{\degree}$. However, with the appearance of the second sharp first-order peak, the transition, as well as its extend at $T_{c}$ change and the NC state (i.e., $\Delta C=0$) is recovered at temperatures right above the peak [Fig. \ref{Fig_4}(i)]. 

For off-alignments beyond $\alpha=\SI{2.7}{\degree}$, the thermal-relaxation curves indicate non-equilibrium behavior and a determination of the specific heat is not possible anymore. As shown in Fig. \ref{Fig_4}(o), the relaxation curves for $\alpha$ between \SI{2.93}{\degree} and \SI{3.32}{\degree} show a sudden jump-like increase in temperature, corresponding to an abrupt release of latent heat of up to $\SI{0.16}{\joule\mole^{-1}}$ at $\alpha=\SI{3.32}{\degree}$. With increasing $\alpha$, the jump continuously shifts to lower temperatures, whereas its height increases.  Finally, for $\alpha\ge\SI{3.5}{\degree}$, the NC state is recovered and the thermal relaxation becomes featureless. We observed qualitatively similar behavior for negative rotation angles and at $\SI{21.7}{\tesla}$, which confirms the intrinsic nature of these non-equilibrium heat releases \cite{NoteSuppl}.

In contrast to the peak at $T_{high}$, the signature at $T_{low}$ decouples from the SC transition with increasing $\alpha$. This indicates a further transition within the high-field SC state. In order to examine if this transition is unique to the FFLO state, we measured the specific heat for a slight off-alignment ($\alpha=\SI{1.4}{\degree}$) at various fields [Fig. \ref{Fig_2}(b)]. We found that the sharp signature, which occurs a few hundred \SI{}{\milli\kelvin} below $T_{c}$, persist down to at least \SI{6}{\tesla}, proving that it is not unique to the high-field phase.

\textcolor{black}{The SC phase boundary, calculated under the assumption of an isotropic in-plane Fermi velocity $v_F$ \cite{Wosnitza2018} and s-wave pairing [dashed line in Fig. \ref{Fig_1}(b)], yields an excellent agreement with the phase boundary determined by means of specific heat. Accordingly, the FFLO state stabilizes below the tricritical point at $T^{*}=0.56 T_\mathrm{c}\approx\SI{5.1}{\kelvin}$ and above $\mu_0H_\mathrm{c}(T^*)\approx\SI{19.3}{\tesla}$.}
\begin{figure}[tb]
	\includegraphics{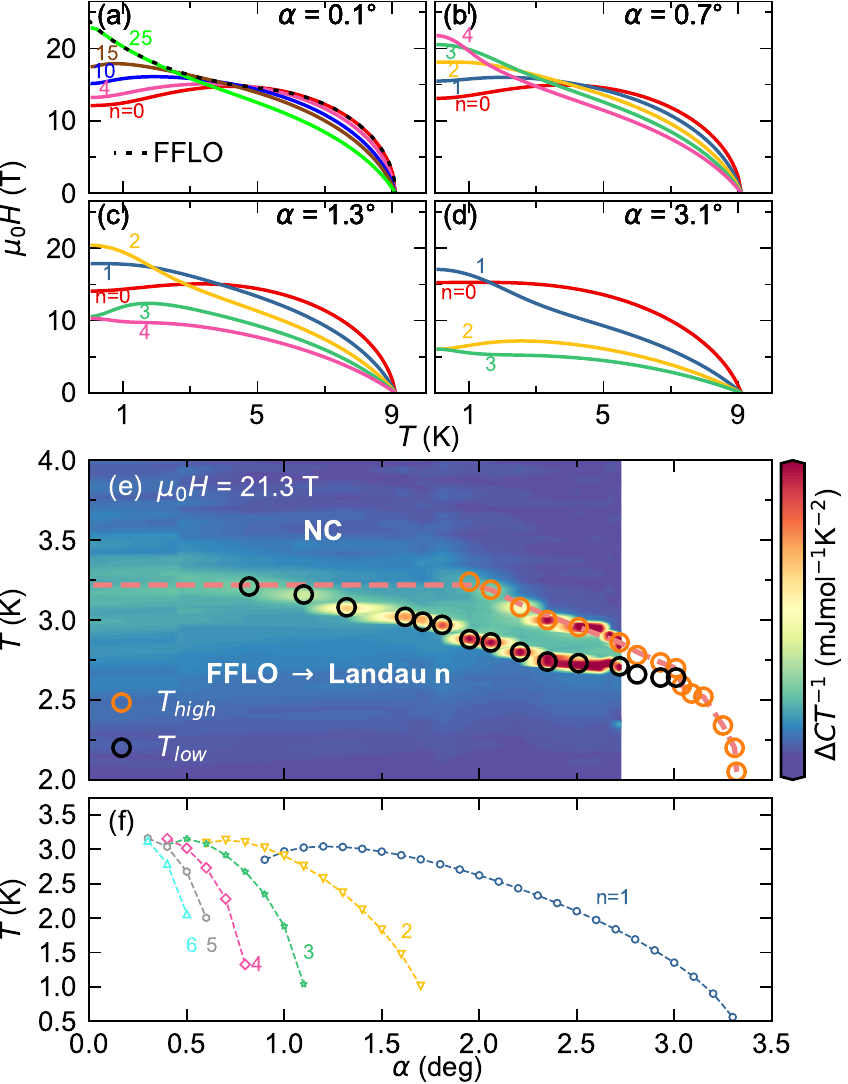}
	\caption{\label{Fig_5}\textcolor{black}{(a)-(d) The upper critical fields calculated according to Eq. (\ref{Eq_selfcons}) for different $\alpha$ and $n$. For $\alpha=\SI{0.1}{\degree}$, the FFLO solution is plotted as dashed line for comparison. (e) $\Delta C/T$ of $\kappa$-(ET)$_2$Cu(NCS)$_2$ at \SI{21.3}{\tesla} as function of $T$ and $\alpha$. Below $\alpha=\SI{2.7}{\degree}$, $T_{low}$ and $T_{high}$ correspond to the position of the peaks seen in specific heat while they are directly determined by the discontinuities in the relaxation curve for larger $\alpha$ [Fig. \ref{Fig_4}(o)]. (f) Calculated critical temperature at \SI{21.3}{\tesla} as function of $\alpha$ for selected Landau states $n$. The envelope of the curves marks the NC-SC transition.}}
\end{figure}

\textcolor{black}{In a next step}, we discuss the suppression of the FFLO phase with increasing tilt angle $\alpha$. As the main result of our work, Fig. \ref{Fig_5}(e) provides a $T$-$\alpha$ phase diagram of the SC state in $\kappa$-(ET)$_2$Cu(NCS)$_2$ at \SI{21.3}{\tesla}. We show a contour plot of $\Delta C/T$, interpolated from the data presented in Fig. \ref{Fig_4}. Additionally, the first-order transitions at $T_{low}$ and $T_{high}$ are marked by black and orange circles, respectively. 

Three observations are noteworthy: (i) The SC transition temperature is nearly unaffected up to $\alpha=\SI{2}{\degree}$. This is in stark contrast to $\beta''$-(ET)$_2$SF$_5$CH$_2$CF$_2$SO$_3$, for which the FFLO phase is already suppressed for $\alpha>\SI{0.5}{\degree}$ \cite{Beyer2012}. (ii) A second high-field SC phase emerges between $\alpha =\SI{0.8}{\degree}$ and \SI{3}{\degree}. (iii) The SC transition crosses over from a second-order behavior for $\alpha<\SI{2}{\degree}$ to a pronounced first-order one, before superconductivity is finally suppressed around \SI{3.5}{\degree}. \textcolor{black}{The last observation explains the controversy that some studies claim the NC-FFLO to be of second order \cite{Agosta2017,Tsuchiya2015} while other found evidence for a first-order transition \cite{Lortz2007}.}

Theoretical considerations of the SC phase diagram for Q2D superconductors in the presence of paramagnetic as well as orbital effects predict that, below $T^*$, the Abrikosov solution \cite{Abrikosov1957} for the SC order parameter becomes unstable against the formation of multi-quanta states of higher Landau indices $n$ \cite{Bulaevskii1973,Shimahara1997,Buzdin1996,Houzet2000,Houzet2002}. Hence, for increasing orbital contributions (i.e., increasing $\alpha$), the FFLO order parameter (corresponding to $n\rightarrow\infty$) is expected to transform into the Abrikosov lattice ($n=0$) via intermediate states of higher $n$. For BCS superconductors in the weak-coupling limit, the linearized self-consistency equation, describing the upper critical field as function of $T$, $n$, and $\alpha$, is given in Ref. \cite{Shimahara1997}, Eq. (2.9). Introducing the characteristic fields
\begin{equation}
\bar{H}_Z = \frac{\pi k_B T_c}{\mu_B}\text{ ; }
\bar{H}_{orb} = \frac{\Phi_0}{2\pi}\left(\frac{\pi k_B T_c}{\hbar v_F}\right)^2\mathrm{ ,}\label{Eq_Horb}
\end{equation}
and recognizing the Laguerre polynomial $L_n$, the self-consistency equation can be simplified to
\begin{widetext}
\begin{equation}\textcolor{black}{
-\log \left(\frac{T}{T_c}\right)=
\int_0^\infty\frac{du}{\sinh u}\left\lbrace 1-\cos\left(\frac{H}{\bar{H}_Z}\frac{T_c}{T}u\right) \left[e^{-\frac{1}{16}u^2\frac{H}{\bar{H}_{orb}}\sin\alpha\left(\frac{T_c}{T}\right)^2}
L_n\left(\frac{1}{8}u^2\frac{H}{\bar{H}_{orb}}\sin\alpha\left(\frac{T_c}{T}\right)^2\right)\right]\right\rbrace\mathrm{ .}\label{Eq_selfcons}}
\end{equation}
\end{widetext}
Thereby, $k_B$ is the Boltzmann constant, $\hbar$ the reduced Planck constant, $\mu_B$ the Bohr magneton, and $\Phi_0$ the magnetic flux quantum \cite{NoteSuppl}.

By solving Eq. (\ref{Eq_selfcons}) numerically, we obtain the phase diagram of $\kappa$-ET$_2$Cu(NCS)$_2$ for various angles $\alpha$. We focus on the lowest Landau levels, since those are stabilized by increased off-alignment. \textcolor{black}{However, the solution of Eq. (\ref{Eq_selfcons}) converges to the FFLO result for high $n$ and small $\alpha$ \cite{Shimahara1997,Houzet2000,NoteSuppl}}. We estimate $\bar{H}_{orb}$ from the initial slope $\left.\frac{dH_{c2}}{dT}\right|_{T=T_c}=-\SI{0.66}{\tesla\per\kelvin}$ for fields applied perpendicular to the SC layers \cite{Lortz2007,NoteSuppl}. 

The calculated $H$-$T$-phase boundaries for the $n$th Landau states are shown in Figs. \ref{Fig_5} (a)-(d) for selected angles $\alpha$ \cite{NoteSuppl}. The ground state at a given $T$ and $\alpha$ is provided by the state with the highest critical field. Compared to Fig. \ref{Fig_1}(b), the weak-coupling calculation [Eq. (\ref{Eq_selfcons})] underestimates the phase boundaries. Indeed, $\kappa$-(ET)$_2$Cu(NCS)$_2$ is a strong-coupling superconductor \cite{Mueller2002,Wosnitza2003}. The jump in the specific heat at zero field $\Delta C\gamma^{-1} T_c^{-1}=2.3$ is enhanced by a factor of $\eta_{H_c}^2=1.6$ compared to the universal BCS value of \SI{1.43}{}. Similar strong-coupling enhancement factors of the thermodynamic critical field $\eta_{H_c}$ are found in the well-studied strong-coupling superconductors Hg and Pb in which they lead to a temperature-dependent enhancement of the orbital critical field by a factor of \SI{1.2}{}-\SI{1.3}{} \cite{Rainer1974}. Dividing the calculated FFLO phase boundary of $\kappa$-ET$_2$Cu(NCS)$_2$ [dashed line in Fig. \ref{Fig_1}(b)] by \SI{1.32}{}, leads to a perfect agreement with the phase boundaries calculated from Eq. (\ref{Eq_selfcons}) in the high-$n$ limit [Fig. \ref{Fig_5}(a)]. 

Figure \ref{Fig_5}(f) shows the critical temperatures of the first Landau levels as function of $\alpha$ at about \SI{21.3}{\tesla}. We multiplied these data accordingly by \SI{1.32}{} in order to compare them to the field studied in Fig. \ref{Fig_5}(e). The curves reproduce our main experimental results: (i) The pure FFLO state is rapidly suppressed by tilting the field away from the SC planes. However, with increasing angle up to $\alpha\approx\SI{0.7}{\degree}$, the FFLO state is replaced by a cascade of closely lying higher-order Landau states [Fig. \ref{Fig_5}(b)], which prevent a suppression of $T_c$ by the increasing orbital effects. It is likely that the strong fluctuations in $\kappa$-ET$_2$Cu(NCS)$_2$ \cite{Lortz2007,Sasaki2002,Tsuchiya2015} mask individual transitions. (ii) For intermediate off-alignment ($\alpha\ge\SI{1.3}{\degree}$), only the first two Landau levels exceed the Abrikosov state and their transitions are well separated [Fig. \ref{Fig_5}(c)]. \textcolor{black}{Considering the Ginzburg-Landau functional close to $T^*$, it has been predicted that the SC-NC transition may turn first order for low $T$ and $n$, if orbital effects are increased \cite{Houzet2000,Houzet2006}. This is confirmed by the first-order signature found at $T_{high}$ in the specific-heat data. }(iii) Once $\alpha$ exceeds \SI{1.8}{\degree}, the increase of orbital effects cannot be compensated anymore by a structural adaptation of the SC order parameter. Hence, $T_c$ is strongly reduced by a further tilt of the field. Finally, the high-field SC state is suppressed at $\alpha = \SI{3.3}{\degree}$, above which no solution for Eq. (\ref{Eq_selfcons}) exists. 
\textcolor{black}{An underlying d-wave symmetry of the order parameter would not change qualitatively the angular dependence of $T_c$ and higher-order Landau states are still required to explain the observed angular stability \cite{NoteSuppl}.}
	
\textcolor{black}{At first, it seems possible that the feature at $T_{low}$ marks a second transition between Landau states. Since it persist down to low fields \cite{Bergk2011}, an overlayed vortex-melting transition is more likely \cite{Mola2001,Sasaki2002}. For perpendicular field geometry, vortex melting in $\kappa$-ET$_2$Cu(NCS)$_2$ has been studied extensively \cite{Mola2001,Sasaki2002,Konoike2009,Yin2007}. However, the $T_{low}$ feature is not essential for our main finding, namely the angular-dependent transition between FFLO and Abrikosov-like states.}

Details of the $T$-$\alpha$ phase diagram might be further altered by (i) strong-coupling effects \cite{Ashauer1987}, (ii) interlayer coupling \cite{Croitoru2017}, (iii) fluctuations \cite{Shimahara1998}, \textcolor{black}{(iv) spin-orbit coupling \cite{Zwicknagl2017}} and, (v) to a smaller extend, by vortex interactions \cite{Brandt1995}. \textcolor{black}{However, we would like to emphasize that none of the aforementioned effects can explain the observed angular stability of the high-field SC state \cite{NoteSuppl}. Instead, a modification of the FFLO order parameter towards higher-order Landau states is required. }

In summary, we presented angular-dependent specific-heat measurements of \mbox{$\kappa$-(ET)$_2$Cu(NCS)$_2$} focusing on the suppression of the high-field FFLO state by orbital effects. In particular, we found a large angular stability of the high-field SC phase up to a few degrees, a change of the nature of the SC transition towards first order, as well as the appearance of a second SC state with increasing orbital contributions. The results indicate that the pure FFLO order parameter in $\kappa$-(ET)$_2$Cu(NCS)$_2$ transforms into higher-order Landau states with increasing orbital contribution. Angular-resolved measurements on Q2D superconductors, hence, offer the unique opportunity to systematically study new manifestations of superconductivity above the Pauli limit beyond the pure FFLO description. 

We acknowledge the support of the HLD at HZDR, member of the European Magnetic Field Laboratory (EMFL) and by the DFG through the W\"{u}rzburg-Dresden Cluster of Excellence on 
Complexity and Topology in Quantum Matter---$ct.qmat$ (EXC 2147, Project No.\ 390858490) and the ANR-DFG grant Fermi-NESt. JAS acknowledges support from the Independent Research/Development program while serving at the National Science Foundation. Fruitful discussions with T. Gottschall and J. Gronemann are highly appreciated. 
\bibliography{20220313_kappa_ET_bibtex}

\begin{thebibliography}{57}%
\makeatletter
\providecommand \@ifxundefined [1]{%
 \@ifx{#1\undefined}
}%
\providecommand \@ifnum [1]{%
 \ifnum #1\expandafter \@firstoftwo
 \else \expandafter \@secondoftwo
 \fi
}%
\providecommand \@ifx [1]{%
 \ifx #1\expandafter \@firstoftwo
 \else \expandafter \@secondoftwo
 \fi
}%
\providecommand \natexlab [1]{#1}%
\providecommand \enquote  [1]{``#1''}%
\providecommand \bibnamefont  [1]{#1}%
\providecommand \bibfnamefont [1]{#1}%
\providecommand \citenamefont [1]{#1}%
\providecommand \href@noop [0]{\@secondoftwo}%
\providecommand \href [0]{\begingroup \@sanitize@url \@href}%
\providecommand \@href[1]{\@@startlink{#1}\@@href}%
\providecommand \@@href[1]{\endgroup#1\@@endlink}%
\providecommand \@sanitize@url [0]{\catcode `\\12\catcode `\$12\catcode
  `\&12\catcode `\#12\catcode `\^12\catcode `\_12\catcode `\%12\relax}%
\providecommand \@@startlink[1]{}%
\providecommand \@@endlink[0]{}%
\providecommand \url  [0]{\begingroup\@sanitize@url \@url }%
\providecommand \@url [1]{\endgroup\@href {#1}{\urlprefix }}%
\providecommand \urlprefix  [0]{URL }%
\providecommand \Eprint [0]{\href }%
\providecommand \doibase [0]{https://doi.org/}%
\providecommand \selectlanguage [0]{\@gobble}%
\providecommand \bibinfo  [0]{\@secondoftwo}%
\providecommand \bibfield  [0]{\@secondoftwo}%
\providecommand \translation [1]{[#1]}%
\providecommand \BibitemOpen [0]{}%
\providecommand \bibitemStop [0]{}%
\providecommand \bibitemNoStop [0]{.\EOS\space}%
\providecommand \EOS [0]{\spacefactor3000\relax}%
\providecommand \BibitemShut  [1]{\csname bibitem#1\endcsname}%
\let\auto@bib@innerbib\@empty
\bibitem [{\citenamefont {Tinkham}(2004)}]{tinkham2004}%
  \BibitemOpen
  \bibfield  {author} {\bibinfo {author} {\bibfnamefont {M.}~\bibnamefont
  {Tinkham}},\ }\href {https://books.google.de/books?id=VpUk3NfwDIkC} {\emph
  {\bibinfo {title} {Introduction to Superconductivity}}},\ Dover Books on
  Physics Series\ (\bibinfo  {publisher} {Dover Publications},\ \bibinfo {year}
  {2004})\BibitemShut {NoStop}%
\bibitem [{\citenamefont {Chandrasekhar}(1962)}]{Chandrasekhar1962}%
  \BibitemOpen
  \bibfield  {author} {\bibinfo {author} {\bibfnamefont {B.~S.}\ \bibnamefont
  {Chandrasekhar}},\ }\href {https://doi.org/10.1063/1.1777362} {\bibfield
  {journal} {\bibinfo  {journal} {Appl. Phys. Lett.}\ }\textbf {\bibinfo
  {volume} {1}},\ \bibinfo {pages} {7} (\bibinfo {year} {1962})}\BibitemShut
  {NoStop}%
\bibitem [{\citenamefont {Clogston}(1962)}]{Clogston1962}%
  \BibitemOpen
  \bibfield  {author} {\bibinfo {author} {\bibfnamefont {A.~M.}\ \bibnamefont
  {Clogston}},\ }\href {https://doi.org/10.1103/PhysRevLett.9.266} {\bibfield
  {journal} {\bibinfo  {journal} {Phys. Rev. Lett.}\ }\textbf {\bibinfo
  {volume} {9}},\ \bibinfo {pages} {266} (\bibinfo {year} {1962})}\BibitemShut
  {NoStop}%
\bibitem [{\citenamefont {Werthamer}\ \emph {et~al.}(1966)\citenamefont
  {Werthamer}, \citenamefont {Helfand},\ and\ \citenamefont
  {Hohenberg}}]{Werthamer1966}%
  \BibitemOpen
  \bibfield  {author} {\bibinfo {author} {\bibfnamefont {N.~R.}\ \bibnamefont
  {Werthamer}}, \bibinfo {author} {\bibfnamefont {E.}~\bibnamefont {Helfand}},\
  and\ \bibinfo {author} {\bibfnamefont {P.~C.}\ \bibnamefont {Hohenberg}},\
  }\href {https://doi.org/10.1103/PhysRev.147.295} {\bibfield  {journal}
  {\bibinfo  {journal} {Phys. Rev.}\ }\textbf {\bibinfo {volume} {147}},\
  \bibinfo {pages} {295} (\bibinfo {year} {1966})}\BibitemShut {NoStop}%
\bibitem [{\citenamefont {Matsuda}\ and\ \citenamefont
  {Shimahara}(2007)}]{Matsuda2007}%
  \BibitemOpen
  \bibfield  {author} {\bibinfo {author} {\bibfnamefont {Y.}~\bibnamefont
  {Matsuda}}\ and\ \bibinfo {author} {\bibfnamefont {H.}~\bibnamefont
  {Shimahara}},\ }\href {https://doi.org/10.1143/JPSJ.76.051005} {\bibfield
  {journal} {\bibinfo  {journal} {J. Phys. Soc. Jpn.}\ }\textbf {\bibinfo
  {volume} {76}},\ \bibinfo {pages} {051005} (\bibinfo {year}
  {2007})}\BibitemShut {NoStop}%
\bibitem [{\citenamefont {Zwicknagl}\ and\ \citenamefont
  {Wosnitza}(2010)}]{Zwicknagl2010}%
  \BibitemOpen
  \bibfield  {author} {\bibinfo {author} {\bibfnamefont {G.}~\bibnamefont
  {Zwicknagl}}\ and\ \bibinfo {author} {\bibfnamefont {J.}~\bibnamefont
  {Wosnitza}},\ }\href {https://doi.org/10.1142/S0217979210056396} {\bibfield
  {journal} {\bibinfo  {journal} {Int. J. Mod. Phys. B}\ }\textbf {\bibinfo
  {volume} {24}},\ \bibinfo {pages} {3915} (\bibinfo {year}
  {2010})}\BibitemShut {NoStop}%
\bibitem [{\citenamefont {Fulde}\ and\ \citenamefont
  {Ferrell}(1964)}]{Fulde1964}%
  \BibitemOpen
  \bibfield  {author} {\bibinfo {author} {\bibfnamefont {P.}~\bibnamefont
  {Fulde}}\ and\ \bibinfo {author} {\bibfnamefont {R.~A.}\ \bibnamefont
  {Ferrell}},\ }\href {https://doi.org/10.1103/PhysRev.135.A550} {\bibfield
  {journal} {\bibinfo  {journal} {Phys. Rev.}\ }\textbf {\bibinfo {volume}
  {135}},\ \bibinfo {pages} {A550} (\bibinfo {year} {1964})}\BibitemShut
  {NoStop}%
\bibitem [{\citenamefont {Larkin}\ and\ \citenamefont
  {Ovchinnikov}(1964)}]{Larkin1964}%
  \BibitemOpen
  \bibfield  {author} {\bibinfo {author} {\bibfnamefont {A.~I.}\ \bibnamefont
  {Larkin}}\ and\ \bibinfo {author} {\bibfnamefont {Y.~N.}\ \bibnamefont
  {Ovchinnikov}},\ }\href@noop {} {\bibfield  {journal} {\bibinfo  {journal}
  {Zh. Eksp. Teor. Fiz.}\ }\textbf {\bibinfo {volume} {47}},\ \bibinfo {pages}
  {1136} (\bibinfo {year} {1964})}\BibitemShut {NoStop}%
\bibitem [{\citenamefont {Shimahara}\ and\ \citenamefont
  {Rainer}(1997)}]{Shimahara1997}%
  \BibitemOpen
  \bibfield  {author} {\bibinfo {author} {\bibfnamefont {H.}~\bibnamefont
  {Shimahara}}\ and\ \bibinfo {author} {\bibfnamefont {D.}~\bibnamefont
  {Rainer}},\ }\href {https://doi.org/10.1143/JPSJ.66.3591} {\bibfield
  {journal} {\bibinfo  {journal} {J. Phys. Soc. Jpn.}\ }\textbf {\bibinfo
  {volume} {66}},\ \bibinfo {pages} {3591} (\bibinfo {year}
  {1997})}\BibitemShut {NoStop}%
\bibitem [{\citenamefont {Takada}(1970)}]{Takada1970}%
  \BibitemOpen
  \bibfield  {author} {\bibinfo {author} {\bibfnamefont {S.}~\bibnamefont
  {Takada}},\ }\href {https://doi.org/10.1143/PTP.43.27} {\bibfield  {journal}
  {\bibinfo  {journal} {Prog. Theor. Phys.}\ }\textbf {\bibinfo {volume}
  {43}},\ \bibinfo {pages} {27} (\bibinfo {year} {1970})}\BibitemShut {NoStop}%
\bibitem [{\citenamefont {Gruenberg}\ and\ \citenamefont
  {Gunther}(1966)}]{Gruenberg1966}%
  \BibitemOpen
  \bibfield  {author} {\bibinfo {author} {\bibfnamefont {L.~W.}\ \bibnamefont
  {Gruenberg}}\ and\ \bibinfo {author} {\bibfnamefont {L.}~\bibnamefont
  {Gunther}},\ }\href {https://doi.org/10.1103/PhysRevLett.16.996} {\bibfield
  {journal} {\bibinfo  {journal} {Phys. Rev. Lett.}\ }\textbf {\bibinfo
  {volume} {16}},\ \bibinfo {pages} {996} (\bibinfo {year} {1966})}\BibitemShut
  {NoStop}%
\bibitem [{\citenamefont {Brown}(2015)}]{Brown2015}%
  \BibitemOpen
  \bibfield  {author} {\bibinfo {author} {\bibfnamefont {S.~E.}\ \bibnamefont
  {Brown}},\ }\href {https://doi.org/10.1016/j.physc.2015.02.030} {\bibfield
  {journal} {\bibinfo  {journal} {Physica C}\ }\textbf {\bibinfo {volume}
  {514}},\ \bibinfo {pages} {279} (\bibinfo {year} {2015})}\BibitemShut
  {NoStop}%
\bibitem [{\citenamefont {Wosnitza}(2007)}]{Wosnitza2007}%
  \BibitemOpen
  \bibfield  {author} {\bibinfo {author} {\bibfnamefont {J.}~\bibnamefont
  {Wosnitza}},\ }\href {https://doi.org/10.1007/s10909-006-9282-9} {\bibfield
  {journal} {\bibinfo  {journal} {J. Low Temp. Phys.}\ }\textbf {\bibinfo
  {volume} {146}},\ \bibinfo {pages} {641} (\bibinfo {year}
  {2007})}\BibitemShut {NoStop}%
\bibitem [{\citenamefont {Beyer}\ and\ \citenamefont
  {Wosnitza}(2013)}]{Beyer2013a}%
  \BibitemOpen
  \bibfield  {author} {\bibinfo {author} {\bibfnamefont {R.}~\bibnamefont
  {Beyer}}\ and\ \bibinfo {author} {\bibfnamefont {J.}~\bibnamefont
  {Wosnitza}},\ }\href {https://doi.org/10.1063/1.4794996} {\bibfield
  {journal} {\bibinfo  {journal} {Low Temp. Phys.}\ }\textbf {\bibinfo {volume}
  {39}},\ \bibinfo {pages} {225} (\bibinfo {year} {2013})}\BibitemShut
  {NoStop}%
\bibitem [{\citenamefont {Wosnitza}(2018)}]{Wosnitza2018}%
  \BibitemOpen
  \bibfield  {author} {\bibinfo {author} {\bibfnamefont {J.}~\bibnamefont
  {Wosnitza}},\ }\href {https://doi.org/https://doi.org/10.1002/andp.201700282}
  {\bibfield  {journal} {\bibinfo  {journal} {Ann. Phys.}\ }\textbf {\bibinfo
  {volume} {530}},\ \bibinfo {pages} {1700282} (\bibinfo {year}
  {2018})}\BibitemShut {NoStop}%
\bibitem [{\citenamefont {Lortz}\ \emph {et~al.}(2007)\citenamefont {Lortz},
  \citenamefont {Wang}, \citenamefont {Demuer}, \citenamefont {B\"ottger},
  \citenamefont {Bergk}, \citenamefont {Zwicknagl}, \citenamefont {Nakazawa},\
  and\ \citenamefont {Wosnitza}}]{Lortz2007}%
  \BibitemOpen
  \bibfield  {author} {\bibinfo {author} {\bibfnamefont {R.}~\bibnamefont
  {Lortz}}, \bibinfo {author} {\bibfnamefont {Y.}~\bibnamefont {Wang}},
  \bibinfo {author} {\bibfnamefont {A.}~\bibnamefont {Demuer}}, \bibinfo
  {author} {\bibfnamefont {P.~H.~M.}\ \bibnamefont {B\"ottger}}, \bibinfo
  {author} {\bibfnamefont {B.}~\bibnamefont {Bergk}}, \bibinfo {author}
  {\bibfnamefont {G.}~\bibnamefont {Zwicknagl}}, \bibinfo {author}
  {\bibfnamefont {Y.}~\bibnamefont {Nakazawa}},\ and\ \bibinfo {author}
  {\bibfnamefont {J.}~\bibnamefont {Wosnitza}},\ }\href
  {https://doi.org/10.1103/PhysRevLett.99.187002} {\bibfield  {journal}
  {\bibinfo  {journal} {Phys. Rev. Lett.}\ }\textbf {\bibinfo {volume} {99}},\
  \bibinfo {pages} {187002} (\bibinfo {year} {2007})}\BibitemShut {NoStop}%
\bibitem [{\citenamefont {Beyer}\ \emph {et~al.}(2012)\citenamefont {Beyer},
  \citenamefont {Bergk}, \citenamefont {Yasin}, \citenamefont {Schlueter},\
  and\ \citenamefont {Wosnitza}}]{Beyer2012}%
  \BibitemOpen
  \bibfield  {author} {\bibinfo {author} {\bibfnamefont {R.}~\bibnamefont
  {Beyer}}, \bibinfo {author} {\bibfnamefont {B.}~\bibnamefont {Bergk}},
  \bibinfo {author} {\bibfnamefont {S.}~\bibnamefont {Yasin}}, \bibinfo
  {author} {\bibfnamefont {J.~A.}\ \bibnamefont {Schlueter}},\ and\ \bibinfo
  {author} {\bibfnamefont {J.}~\bibnamefont {Wosnitza}},\ }\href
  {https://doi.org/10.1103/PhysRevLett.109.027003} {\bibfield  {journal}
  {\bibinfo  {journal} {Phys. Rev. Lett.}\ }\textbf {\bibinfo {volume} {109}},\
  \bibinfo {pages} {027003} (\bibinfo {year} {2012})}\BibitemShut {NoStop}%
\bibitem [{\citenamefont {Agosta}\ \emph {et~al.}(2017)\citenamefont {Agosta},
  \citenamefont {Fortune}, \citenamefont {Hannahs}, \citenamefont {Gu},
  \citenamefont {Liang}, \citenamefont {Park},\ and\ \citenamefont
  {Schlueter}}]{Agosta2017}%
  \BibitemOpen
  \bibfield  {author} {\bibinfo {author} {\bibfnamefont {C.~C.}\ \bibnamefont
  {Agosta}}, \bibinfo {author} {\bibfnamefont {N.~A.}\ \bibnamefont {Fortune}},
  \bibinfo {author} {\bibfnamefont {S.~T.}\ \bibnamefont {Hannahs}}, \bibinfo
  {author} {\bibfnamefont {S.}~\bibnamefont {Gu}}, \bibinfo {author}
  {\bibfnamefont {L.}~\bibnamefont {Liang}}, \bibinfo {author} {\bibfnamefont
  {J.-H.}\ \bibnamefont {Park}},\ and\ \bibinfo {author} {\bibfnamefont
  {J.~A.}\ \bibnamefont {Schlueter}},\ }\href
  {https://doi.org/10.1103/PhysRevLett.118.267001} {\bibfield  {journal}
  {\bibinfo  {journal} {Phys. Rev. Lett.}\ }\textbf {\bibinfo {volume} {118}},\
  \bibinfo {pages} {267001} (\bibinfo {year} {2017})}\BibitemShut {NoStop}%
\bibitem [{\citenamefont {Agosta}\ \emph {et~al.}(2012)\citenamefont {Agosta},
  \citenamefont {Jin}, \citenamefont {Coniglio}, \citenamefont {Smith},
  \citenamefont {Cho}, \citenamefont {Stroe}, \citenamefont {Martin},
  \citenamefont {Tozer}, \citenamefont {Murphy}, \citenamefont {Palm},
  \citenamefont {Schlueter},\ and\ \citenamefont {Kurmoo}}]{Agosta2012}%
  \BibitemOpen
  \bibfield  {author} {\bibinfo {author} {\bibfnamefont {C.~C.}\ \bibnamefont
  {Agosta}}, \bibinfo {author} {\bibfnamefont {J.}~\bibnamefont {Jin}},
  \bibinfo {author} {\bibfnamefont {W.~A.}\ \bibnamefont {Coniglio}}, \bibinfo
  {author} {\bibfnamefont {B.~E.}\ \bibnamefont {Smith}}, \bibinfo {author}
  {\bibfnamefont {K.}~\bibnamefont {Cho}}, \bibinfo {author} {\bibfnamefont
  {I.}~\bibnamefont {Stroe}}, \bibinfo {author} {\bibfnamefont
  {C.}~\bibnamefont {Martin}}, \bibinfo {author} {\bibfnamefont {S.~W.}\
  \bibnamefont {Tozer}}, \bibinfo {author} {\bibfnamefont {T.~P.}\ \bibnamefont
  {Murphy}}, \bibinfo {author} {\bibfnamefont {E.~C.}\ \bibnamefont {Palm}},
  \bibinfo {author} {\bibfnamefont {J.~A.}\ \bibnamefont {Schlueter}},\ and\
  \bibinfo {author} {\bibfnamefont {M.}~\bibnamefont {Kurmoo}},\ }\href
  {https://doi.org/10.1103/PhysRevB.85.214514} {\bibfield  {journal} {\bibinfo
  {journal} {Phys. Rev. B}\ }\textbf {\bibinfo {volume} {85}},\ \bibinfo
  {pages} {214514} (\bibinfo {year} {2012})}\BibitemShut {NoStop}%
\bibitem [{\citenamefont {Bergk}\ \emph {et~al.}(2011)\citenamefont {Bergk},
  \citenamefont {Demuer}, \citenamefont {Sheikin}, \citenamefont {Wang},
  \citenamefont {Wosnitza}, \citenamefont {Nakazawa},\ and\ \citenamefont
  {Lortz}}]{Bergk2011}%
  \BibitemOpen
  \bibfield  {author} {\bibinfo {author} {\bibfnamefont {B.}~\bibnamefont
  {Bergk}}, \bibinfo {author} {\bibfnamefont {A.}~\bibnamefont {Demuer}},
  \bibinfo {author} {\bibfnamefont {I.}~\bibnamefont {Sheikin}}, \bibinfo
  {author} {\bibfnamefont {Y.}~\bibnamefont {Wang}}, \bibinfo {author}
  {\bibfnamefont {J.}~\bibnamefont {Wosnitza}}, \bibinfo {author}
  {\bibfnamefont {Y.}~\bibnamefont {Nakazawa}},\ and\ \bibinfo {author}
  {\bibfnamefont {R.}~\bibnamefont {Lortz}},\ }\href
  {https://doi.org/10.1103/PhysRevB.83.064506} {\bibfield  {journal} {\bibinfo
  {journal} {Phys. Rev. B}\ }\textbf {\bibinfo {volume} {83}},\ \bibinfo
  {pages} {064506} (\bibinfo {year} {2011})}\BibitemShut {NoStop}%
\bibitem [{\citenamefont {Tsuchiya}\ \emph {et~al.}(2015)\citenamefont
  {Tsuchiya}, \citenamefont {Yamada}, \citenamefont {Sugii}, \citenamefont
  {Graf}, \citenamefont {Brooks}, \citenamefont {Terashima},\ and\
  \citenamefont {Uji}}]{Tsuchiya2015}%
  \BibitemOpen
  \bibfield  {author} {\bibinfo {author} {\bibfnamefont {S.}~\bibnamefont
  {Tsuchiya}}, \bibinfo {author} {\bibfnamefont {J.-i.}\ \bibnamefont
  {Yamada}}, \bibinfo {author} {\bibfnamefont {K.}~\bibnamefont {Sugii}},
  \bibinfo {author} {\bibfnamefont {D.}~\bibnamefont {Graf}}, \bibinfo {author}
  {\bibfnamefont {J.~S.}\ \bibnamefont {Brooks}}, \bibinfo {author}
  {\bibfnamefont {T.}~\bibnamefont {Terashima}},\ and\ \bibinfo {author}
  {\bibfnamefont {S.}~\bibnamefont {Uji}},\ }\href
  {https://doi.org/10.7566/JPSJ.84.034703} {\bibfield  {journal} {\bibinfo
  {journal} {J. Phys. Soc. Jpn.}\ }\textbf {\bibinfo {volume} {84}},\ \bibinfo
  {pages} {034703} (\bibinfo {year} {2015})}\BibitemShut {NoStop}%
\bibitem [{\citenamefont {Sugiura}\ \emph {et~al.}(2019)\citenamefont
  {Sugiura}, \citenamefont {Isono}, \citenamefont {Terashima}, \citenamefont
  {Yasuzuka}, \citenamefont {Schlueter},\ and\ \citenamefont
  {Uji}}]{Sugiura2019}%
  \BibitemOpen
  \bibfield  {author} {\bibinfo {author} {\bibfnamefont {S.}~\bibnamefont
  {Sugiura}}, \bibinfo {author} {\bibfnamefont {T.}~\bibnamefont {Isono}},
  \bibinfo {author} {\bibfnamefont {T.}~\bibnamefont {Terashima}}, \bibinfo
  {author} {\bibfnamefont {S.}~\bibnamefont {Yasuzuka}}, \bibinfo {author}
  {\bibfnamefont {J.~A.}\ \bibnamefont {Schlueter}},\ and\ \bibinfo {author}
  {\bibfnamefont {S.}~\bibnamefont {Uji}},\ }\href
  {https://doi.org/10.1038/s41535-019-0147-2} {\bibfield  {journal} {\bibinfo
  {journal} {npj Quantum Mater.}\ }\textbf {\bibinfo {volume} {4}},\ \bibinfo
  {pages} {7} (\bibinfo {year} {2019})}\BibitemShut {NoStop}%
\bibitem [{\citenamefont {Mayaffre}\ \emph {et~al.}(2014)\citenamefont
  {Mayaffre}, \citenamefont {Krämer}, \citenamefont {Horvatić}, \citenamefont
  {Berthier}, \citenamefont {Miyagawa}, \citenamefont {Kanoda},\ and\
  \citenamefont {Mitrović}}]{Mayaffre2014}%
  \BibitemOpen
  \bibfield  {author} {\bibinfo {author} {\bibfnamefont {H.}~\bibnamefont
  {Mayaffre}}, \bibinfo {author} {\bibfnamefont {S.}~\bibnamefont {Krämer}},
  \bibinfo {author} {\bibfnamefont {M.}~\bibnamefont {Horvatić}}, \bibinfo
  {author} {\bibfnamefont {C.}~\bibnamefont {Berthier}}, \bibinfo {author}
  {\bibfnamefont {K.}~\bibnamefont {Miyagawa}}, \bibinfo {author}
  {\bibfnamefont {K.}~\bibnamefont {Kanoda}},\ and\ \bibinfo {author}
  {\bibfnamefont {V.~F.}\ \bibnamefont {Mitrović}},\ }\href
  {https://doi.org/10.1038/nphys3121} {\bibfield  {journal} {\bibinfo
  {journal} {Nat. Phys.}\ }\textbf {\bibinfo {volume} {10}},\ \bibinfo {pages}
  {928} (\bibinfo {year} {2014})}\BibitemShut {NoStop}%
\bibitem [{\citenamefont {Koutroulakis}\ \emph {et~al.}(2016)\citenamefont
  {Koutroulakis}, \citenamefont {K\"uhne}, \citenamefont {Schlueter},
  \citenamefont {Wosnitza},\ and\ \citenamefont {Brown}}]{Koutroulakis2016}%
  \BibitemOpen
  \bibfield  {author} {\bibinfo {author} {\bibfnamefont {G.}~\bibnamefont
  {Koutroulakis}}, \bibinfo {author} {\bibfnamefont {H.}~\bibnamefont
  {K\"uhne}}, \bibinfo {author} {\bibfnamefont {J.~A.}\ \bibnamefont
  {Schlueter}}, \bibinfo {author} {\bibfnamefont {J.}~\bibnamefont
  {Wosnitza}},\ and\ \bibinfo {author} {\bibfnamefont {S.~E.}\ \bibnamefont
  {Brown}},\ }\href {https://doi.org/10.1103/PhysRevLett.116.067003} {\bibfield
   {journal} {\bibinfo  {journal} {Phys. Rev. Lett.}\ }\textbf {\bibinfo
  {volume} {116}},\ \bibinfo {pages} {067003} (\bibinfo {year}
  {2016})}\BibitemShut {NoStop}%
\bibitem [{\citenamefont {Lin}\ \emph {et~al.}(2020)\citenamefont {Lin},
  \citenamefont {Kim}, \citenamefont {Bauer}, \citenamefont {Ronning},
  \citenamefont {Thompson},\ and\ \citenamefont {Movshovich}}]{Lin2020}%
  \BibitemOpen
  \bibfield  {author} {\bibinfo {author} {\bibfnamefont {S.-Z.}\ \bibnamefont
  {Lin}}, \bibinfo {author} {\bibfnamefont {D.~Y.}\ \bibnamefont {Kim}},
  \bibinfo {author} {\bibfnamefont {E.~D.}\ \bibnamefont {Bauer}}, \bibinfo
  {author} {\bibfnamefont {F.}~\bibnamefont {Ronning}}, \bibinfo {author}
  {\bibfnamefont {J.~D.}\ \bibnamefont {Thompson}},\ and\ \bibinfo {author}
  {\bibfnamefont {R.}~\bibnamefont {Movshovich}},\ }\href
  {https://doi.org/10.1103/PhysRevLett.124.217001} {\bibfield  {journal}
  {\bibinfo  {journal} {Phys. Rev. Lett.}\ }\textbf {\bibinfo {volume} {124}},\
  \bibinfo {pages} {217001} (\bibinfo {year} {2020})}\BibitemShut {NoStop}%
\bibitem [{\citenamefont {Kitagawa}\ \emph {et~al.}(2018)\citenamefont
  {Kitagawa}, \citenamefont {Nakamine}, \citenamefont {Ishida}, \citenamefont
  {Jeevan}, \citenamefont {Geibel},\ and\ \citenamefont
  {Steglich}}]{Kitagawa2018}%
  \BibitemOpen
  \bibfield  {author} {\bibinfo {author} {\bibfnamefont {S.}~\bibnamefont
  {Kitagawa}}, \bibinfo {author} {\bibfnamefont {G.}~\bibnamefont {Nakamine}},
  \bibinfo {author} {\bibfnamefont {K.}~\bibnamefont {Ishida}}, \bibinfo
  {author} {\bibfnamefont {H.~S.}\ \bibnamefont {Jeevan}}, \bibinfo {author}
  {\bibfnamefont {C.}~\bibnamefont {Geibel}},\ and\ \bibinfo {author}
  {\bibfnamefont {F.}~\bibnamefont {Steglich}},\ }\href
  {https://doi.org/10.1103/PhysRevLett.121.157004} {\bibfield  {journal}
  {\bibinfo  {journal} {Phys. Rev. Lett.}\ }\textbf {\bibinfo {volume} {121}},\
  \bibinfo {pages} {157004} (\bibinfo {year} {2018})}\BibitemShut {NoStop}%
\bibitem [{\citenamefont {Cho}\ \emph {et~al.}(2017)\citenamefont {Cho},
  \citenamefont {Yang}, \citenamefont {Yuan}, \citenamefont {Shen},
  \citenamefont {Wolf},\ and\ \citenamefont {Lortz}}]{Cho2017}%
  \BibitemOpen
  \bibfield  {author} {\bibinfo {author} {\bibfnamefont {C.-w.}\ \bibnamefont
  {Cho}}, \bibinfo {author} {\bibfnamefont {J.~H.}\ \bibnamefont {Yang}},
  \bibinfo {author} {\bibfnamefont {N.~F.~Q.}\ \bibnamefont {Yuan}}, \bibinfo
  {author} {\bibfnamefont {J.}~\bibnamefont {Shen}}, \bibinfo {author}
  {\bibfnamefont {T.}~\bibnamefont {Wolf}},\ and\ \bibinfo {author}
  {\bibfnamefont {R.}~\bibnamefont {Lortz}},\ }\href
  {https://doi.org/10.1103/PhysRevLett.119.217002} {\bibfield  {journal}
  {\bibinfo  {journal} {Phys. Rev. Lett.}\ }\textbf {\bibinfo {volume} {119}},\
  \bibinfo {pages} {217002} (\bibinfo {year} {2017})}\BibitemShut {NoStop}%
\bibitem [{\citenamefont {Kasahara}\ \emph {et~al.}(2020)\citenamefont
  {Kasahara}, \citenamefont {Sato}, \citenamefont {Licciardello}, \citenamefont
  {\ifmmode~\check{C}\else \v{C}\fi{}ulo}, \citenamefont
  {Arsenijevi\ifmmode~\acute{c}\else \'{c}\fi{}}, \citenamefont {Ottenbros},
  \citenamefont {Tominaga}, \citenamefont {B\"oker}, \citenamefont {Eremin},
  \citenamefont {Shibauchi}, \citenamefont {Wosnitza}, \citenamefont {Hussey},\
  and\ \citenamefont {Matsuda}}]{Kasahara2020}%
  \BibitemOpen
  \bibfield  {author} {\bibinfo {author} {\bibfnamefont {S.}~\bibnamefont
  {Kasahara}}, \bibinfo {author} {\bibfnamefont {Y.}~\bibnamefont {Sato}},
  \bibinfo {author} {\bibfnamefont {S.}~\bibnamefont {Licciardello}}, \bibinfo
  {author} {\bibfnamefont {M.}~\bibnamefont {\ifmmode~\check{C}\else
  \v{C}\fi{}ulo}}, \bibinfo {author} {\bibfnamefont {S.}~\bibnamefont
  {Arsenijevi\ifmmode~\acute{c}\else \'{c}\fi{}}}, \bibinfo {author}
  {\bibfnamefont {T.}~\bibnamefont {Ottenbros}}, \bibinfo {author}
  {\bibfnamefont {T.}~\bibnamefont {Tominaga}}, \bibinfo {author}
  {\bibfnamefont {J.}~\bibnamefont {B\"oker}}, \bibinfo {author} {\bibfnamefont
  {I.}~\bibnamefont {Eremin}}, \bibinfo {author} {\bibfnamefont
  {T.}~\bibnamefont {Shibauchi}}, \bibinfo {author} {\bibfnamefont
  {J.}~\bibnamefont {Wosnitza}}, \bibinfo {author} {\bibfnamefont {N.~E.}\
  \bibnamefont {Hussey}},\ and\ \bibinfo {author} {\bibfnamefont
  {Y.}~\bibnamefont {Matsuda}},\ }\href
  {https://doi.org/10.1103/PhysRevLett.124.107001} {\bibfield  {journal}
  {\bibinfo  {journal} {Phys. Rev. Lett.}\ }\textbf {\bibinfo {volume} {124}},\
  \bibinfo {pages} {107001} (\bibinfo {year} {2020})}\BibitemShut {NoStop}%
\bibitem [{\citenamefont {Kasahara}\ \emph {et~al.}(2021)\citenamefont
  {Kasahara}, \citenamefont {Suzuki}, \citenamefont {Machida}, \citenamefont
  {Sato}, \citenamefont {Ukai}, \citenamefont {Murayama}, \citenamefont
  {Suetsugu}, \citenamefont {Kasahara}, \citenamefont {Shibauchi},
  \citenamefont {Hanaguri},\ and\ \citenamefont {Matsuda}}]{Kasahara2021}%
  \BibitemOpen
  \bibfield  {author} {\bibinfo {author} {\bibfnamefont {S.}~\bibnamefont
  {Kasahara}}, \bibinfo {author} {\bibfnamefont {H.}~\bibnamefont {Suzuki}},
  \bibinfo {author} {\bibfnamefont {T.}~\bibnamefont {Machida}}, \bibinfo
  {author} {\bibfnamefont {Y.}~\bibnamefont {Sato}}, \bibinfo {author}
  {\bibfnamefont {Y.}~\bibnamefont {Ukai}}, \bibinfo {author} {\bibfnamefont
  {H.}~\bibnamefont {Murayama}}, \bibinfo {author} {\bibfnamefont
  {S.}~\bibnamefont {Suetsugu}}, \bibinfo {author} {\bibfnamefont
  {Y.}~\bibnamefont {Kasahara}}, \bibinfo {author} {\bibfnamefont
  {T.}~\bibnamefont {Shibauchi}}, \bibinfo {author} {\bibfnamefont
  {T.}~\bibnamefont {Hanaguri}},\ and\ \bibinfo {author} {\bibfnamefont
  {Y.}~\bibnamefont {Matsuda}},\ }\href
  {https://doi.org/10.1103/PhysRevLett.127.257001} {\bibfield  {journal}
  {\bibinfo  {journal} {Phys. Rev. Lett.}\ }\textbf {\bibinfo {volume} {127}},\
  \bibinfo {pages} {257001} (\bibinfo {year} {2021})}\BibitemShut {NoStop}%
\bibitem [{\citenamefont {Cho}\ \emph {et~al.}(2021)\citenamefont {Cho},
  \citenamefont {Lyu}, \citenamefont {Ng}, \citenamefont {He}, \citenamefont
  {Lo}, \citenamefont {Chareev}, \citenamefont {Abdel-Baset}, \citenamefont
  {Abdel-Hafiez},\ and\ \citenamefont {Lortz}}]{Cho2021}%
  \BibitemOpen
  \bibfield  {author} {\bibinfo {author} {\bibfnamefont {C.-w.}\ \bibnamefont
  {Cho}}, \bibinfo {author} {\bibfnamefont {J.}~\bibnamefont {Lyu}}, \bibinfo
  {author} {\bibfnamefont {C.~Y.}\ \bibnamefont {Ng}}, \bibinfo {author}
  {\bibfnamefont {J.~J.}\ \bibnamefont {He}}, \bibinfo {author} {\bibfnamefont
  {K.~T.}\ \bibnamefont {Lo}}, \bibinfo {author} {\bibfnamefont
  {D.}~\bibnamefont {Chareev}}, \bibinfo {author} {\bibfnamefont {T.~A.}\
  \bibnamefont {Abdel-Baset}}, \bibinfo {author} {\bibfnamefont
  {M.}~\bibnamefont {Abdel-Hafiez}},\ and\ \bibinfo {author} {\bibfnamefont
  {R.}~\bibnamefont {Lortz}},\ }\href
  {https://doi.org/10.1038/s41467-021-23976-2} {\bibfield  {journal} {\bibinfo
  {journal} {Nat. Commun.}\ }\textbf {\bibinfo {volume} {12}},\ \bibinfo
  {pages} {3676} (\bibinfo {year} {2021})}\BibitemShut {NoStop}%
\bibitem [{\citenamefont {Burkhardt}\ and\ \citenamefont
  {Rainer}(1994)}]{Burkhardt1994}%
  \BibitemOpen
  \bibfield  {author} {\bibinfo {author} {\bibfnamefont {H.}~\bibnamefont
  {Burkhardt}}\ and\ \bibinfo {author} {\bibfnamefont {D.}~\bibnamefont
  {Rainer}},\ }\href {https://doi.org/https://doi.org/10.1002/andp.19945060305}
  {\bibfield  {journal} {\bibinfo  {journal} {Ann. Phys.}\ }\textbf {\bibinfo
  {volume} {506}},\ \bibinfo {pages} {181} (\bibinfo {year}
  {1994})}\BibitemShut {NoStop}%
\bibitem [{\citenamefont {Bulaevskii}(1973)}]{Bulaevskii1973}%
  \BibitemOpen
  \bibfield  {author} {\bibinfo {author} {\bibfnamefont {L.}~\bibnamefont
  {Bulaevskii}},\ }\href@noop {} {\bibfield  {journal} {\bibinfo  {journal}
  {Zh. Eksp. Teor. Fiz}\ }\textbf {\bibinfo {volume} {65}},\ \bibinfo {pages}
  {1278} (\bibinfo {year} {1973})}\BibitemShut {NoStop}%
\bibitem [{\citenamefont {Buzdin}\ and\ \citenamefont
  {Brison}(1996)}]{Buzdin1996}%
  \BibitemOpen
  \bibfield  {author} {\bibinfo {author} {\bibfnamefont {A.~I.}\ \bibnamefont
  {Buzdin}}\ and\ \bibinfo {author} {\bibfnamefont {J.~P.}\ \bibnamefont
  {Brison}},\ }\href {https://doi.org/10.1209/epl/i1996-00152-9} {\bibfield
  {journal} {\bibinfo  {journal} {EPL}\ }\textbf {\bibinfo {volume} {35}},\
  \bibinfo {pages} {707} (\bibinfo {year} {1996})}\BibitemShut {NoStop}%
\bibitem [{\citenamefont {Houzet}\ and\ \citenamefont
  {Buzdin}(2000)}]{Houzet2000}%
  \BibitemOpen
  \bibfield  {author} {\bibinfo {author} {\bibfnamefont {M.}~\bibnamefont
  {Houzet}}\ and\ \bibinfo {author} {\bibfnamefont {A.}~\bibnamefont
  {Buzdin}},\ }\href {https://doi.org/10.1209/epl/i2000-00281-7} {\bibfield
  {journal} {\bibinfo  {journal} {EPL}\ }\textbf {\bibinfo {volume} {50}},\
  \bibinfo {pages} {375} (\bibinfo {year} {2000})}\BibitemShut {NoStop}%
\bibitem [{\citenamefont {Wright}\ \emph {et~al.}(2011)\citenamefont {Wright},
  \citenamefont {Green}, \citenamefont {Kuhns}, \citenamefont {Reyes},
  \citenamefont {Brooks}, \citenamefont {Schlueter}, \citenamefont {Kato},
  \citenamefont {Yamamoto}, \citenamefont {Kobayashi},\ and\ \citenamefont
  {Brown}}]{Wright2011}%
  \BibitemOpen
  \bibfield  {author} {\bibinfo {author} {\bibfnamefont {J.~A.}\ \bibnamefont
  {Wright}}, \bibinfo {author} {\bibfnamefont {E.}~\bibnamefont {Green}},
  \bibinfo {author} {\bibfnamefont {P.}~\bibnamefont {Kuhns}}, \bibinfo
  {author} {\bibfnamefont {A.}~\bibnamefont {Reyes}}, \bibinfo {author}
  {\bibfnamefont {J.}~\bibnamefont {Brooks}}, \bibinfo {author} {\bibfnamefont
  {J.}~\bibnamefont {Schlueter}}, \bibinfo {author} {\bibfnamefont
  {R.}~\bibnamefont {Kato}}, \bibinfo {author} {\bibfnamefont {H.}~\bibnamefont
  {Yamamoto}}, \bibinfo {author} {\bibfnamefont {M.}~\bibnamefont
  {Kobayashi}},\ and\ \bibinfo {author} {\bibfnamefont {S.~E.}\ \bibnamefont
  {Brown}},\ }\href {https://doi.org/10.1103/PhysRevLett.107.087002} {\bibfield
   {journal} {\bibinfo  {journal} {Phys. Rev. Lett.}\ }\textbf {\bibinfo
  {volume} {107}},\ \bibinfo {pages} {087002} (\bibinfo {year}
  {2011})}\BibitemShut {NoStop}%
\bibitem [{\citenamefont {Fortune}\ \emph {et~al.}(2018)\citenamefont
  {Fortune}, \citenamefont {Agosta}, \citenamefont {Hannahs},\ and\
  \citenamefont {Schlueter}}]{Fortune2018}%
  \BibitemOpen
  \bibfield  {author} {\bibinfo {author} {\bibfnamefont {N.~A.}\ \bibnamefont
  {Fortune}}, \bibinfo {author} {\bibfnamefont {C.~C.}\ \bibnamefont {Agosta}},
  \bibinfo {author} {\bibfnamefont {S.~T.}\ \bibnamefont {Hannahs}},\ and\
  \bibinfo {author} {\bibfnamefont {J.~A.}\ \bibnamefont {Schlueter}},\ }\href
  {https://doi.org/10.1088/1742-6596/969/1/012072} {\bibfield  {journal}
  {\bibinfo  {journal} {J. Phys.: Conf. Ser.}\ }\textbf {\bibinfo {volume}
  {969}},\ \bibinfo {pages} {012072} (\bibinfo {year} {2018})}\BibitemShut
  {NoStop}%
\bibitem [{\citenamefont {Lebed}(2018)}]{Lebed2018}%
  \BibitemOpen
  \bibfield  {author} {\bibinfo {author} {\bibfnamefont {A.~G.}\ \bibnamefont
  {Lebed}},\ }\href {https://doi.org/10.1103/PhysRevB.97.144504} {\bibfield
  {journal} {\bibinfo  {journal} {Phys. Rev. B}\ }\textbf {\bibinfo {volume}
  {97}},\ \bibinfo {pages} {144504} (\bibinfo {year} {2018})}\BibitemShut
  {NoStop}%
\bibitem [{\citenamefont {Urayama}\ \emph {et~al.}(1988)\citenamefont
  {Urayama}, \citenamefont {Yamochi}, \citenamefont {Saito}, \citenamefont
  {Nozawa}, \citenamefont {Sugano}, \citenamefont {Kinoshita}, \citenamefont
  {Sato}, \citenamefont {Oshima}, \citenamefont {Kawamoto},\ and\ \citenamefont
  {Tanaka}}]{Urayama1988}%
  \BibitemOpen
  \bibfield  {author} {\bibinfo {author} {\bibfnamefont {H.}~\bibnamefont
  {Urayama}}, \bibinfo {author} {\bibfnamefont {H.}~\bibnamefont {Yamochi}},
  \bibinfo {author} {\bibfnamefont {G.}~\bibnamefont {Saito}}, \bibinfo
  {author} {\bibfnamefont {K.}~\bibnamefont {Nozawa}}, \bibinfo {author}
  {\bibfnamefont {T.}~\bibnamefont {Sugano}}, \bibinfo {author} {\bibfnamefont
  {M.}~\bibnamefont {Kinoshita}}, \bibinfo {author} {\bibfnamefont
  {S.}~\bibnamefont {Sato}}, \bibinfo {author} {\bibfnamefont {K.}~\bibnamefont
  {Oshima}}, \bibinfo {author} {\bibfnamefont {A.}~\bibnamefont {Kawamoto}},\
  and\ \bibinfo {author} {\bibfnamefont {J.}~\bibnamefont {Tanaka}},\ }\href
  {https://doi.org/10.1246/cl.1988.55} {\bibfield  {journal} {\bibinfo
  {journal} {Chem. Lett.}\ }\textbf {\bibinfo {volume} {17}},\ \bibinfo {pages}
  {55} (\bibinfo {year} {1988})}\BibitemShut {NoStop}%
\bibitem [{Not()}]{NoteSuppl}%
  \BibitemOpen
  \href@noop {} {}\bibinfo {note} {See Supplemental Materials at (filled out by
  Editor) for further details on the experimental setup, results of
  measurements performed for negative $\alpha$ as well as at 21.7 T and 19.5 T,
  a derivation of Eq. (2), its approximation for large $n$ and small $\alpha$,
  as well as for further information on the numerical calculation of the
  critical fields of the $n$th Landau state. Furthermore, we discuss the
  influence of a d-wave order-parameter symmetry and provide information on
  further effects which might alter the SC phase diagram.}\BibitemShut {Stop}%
\bibitem [{\citenamefont {Wang}\ \emph {et~al.}(2001)\citenamefont {Wang},
  \citenamefont {Plackowski},\ and\ \citenamefont {Junod}}]{Wang2001}%
  \BibitemOpen
  \bibfield  {author} {\bibinfo {author} {\bibfnamefont {Y.}~\bibnamefont
  {Wang}}, \bibinfo {author} {\bibfnamefont {T.}~\bibnamefont {Plackowski}},\
  and\ \bibinfo {author} {\bibfnamefont {A.}~\bibnamefont {Junod}},\ }\href
  {https://doi.org/https://doi.org/10.1016/S0921-4534(01)00617-7} {\bibfield
  {journal} {\bibinfo  {journal} {Physica C}\ }\textbf {\bibinfo {volume}
  {355}},\ \bibinfo {pages} {179 } (\bibinfo {year} {2001})}\BibitemShut
  {NoStop}%
\bibitem [{\citenamefont {Andraka}\ \emph {et~al.}(1989)\citenamefont
  {Andraka}, \citenamefont {Kim}, \citenamefont {Stewart}, \citenamefont
  {Carlson}, \citenamefont {Wang},\ and\ \citenamefont
  {Williams}}]{Andraka1989}%
  \BibitemOpen
  \bibfield  {author} {\bibinfo {author} {\bibfnamefont {B.}~\bibnamefont
  {Andraka}}, \bibinfo {author} {\bibfnamefont {J.~S.}\ \bibnamefont {Kim}},
  \bibinfo {author} {\bibfnamefont {G.~R.}\ \bibnamefont {Stewart}}, \bibinfo
  {author} {\bibfnamefont {K.~D.}\ \bibnamefont {Carlson}}, \bibinfo {author}
  {\bibfnamefont {H.~H.}\ \bibnamefont {Wang}},\ and\ \bibinfo {author}
  {\bibfnamefont {J.~M.}\ \bibnamefont {Williams}},\ }\href
  {https://doi.org/10.1103/PhysRevB.40.11345} {\bibfield  {journal} {\bibinfo
  {journal} {Phys. Rev. B}\ }\textbf {\bibinfo {volume} {40}},\ \bibinfo
  {pages} {11345} (\bibinfo {year} {1989})}\BibitemShut {NoStop}%
\bibitem [{\citenamefont {M\"uller}\ \emph {et~al.}(2002)\citenamefont
  {M\"uller}, \citenamefont {Lang}, \citenamefont {Helfrich}, \citenamefont
  {Steglich},\ and\ \citenamefont {Sasaki}}]{Mueller2002}%
  \BibitemOpen
  \bibfield  {author} {\bibinfo {author} {\bibfnamefont {J.}~\bibnamefont
  {M\"uller}}, \bibinfo {author} {\bibfnamefont {M.}~\bibnamefont {Lang}},
  \bibinfo {author} {\bibfnamefont {R.}~\bibnamefont {Helfrich}}, \bibinfo
  {author} {\bibfnamefont {F.}~\bibnamefont {Steglich}},\ and\ \bibinfo
  {author} {\bibfnamefont {T.}~\bibnamefont {Sasaki}},\ }\href
  {https://doi.org/10.1103/PhysRevB.65.140509} {\bibfield  {journal} {\bibinfo
  {journal} {Phys. Rev. B}\ }\textbf {\bibinfo {volume} {65}},\ \bibinfo
  {pages} {140509} (\bibinfo {year} {2002})}\BibitemShut {NoStop}%
\bibitem [{\citenamefont {Wosnitza}\ \emph {et~al.}(2003)\citenamefont
  {Wosnitza}, \citenamefont {Wanka}, \citenamefont {Hagel}, \citenamefont
  {Reibelt}, \citenamefont {Schweitzer},\ and\ \citenamefont
  {Schlueter}}]{Wosnitza2003}%
  \BibitemOpen
  \bibfield  {author} {\bibinfo {author} {\bibfnamefont {J.}~\bibnamefont
  {Wosnitza}}, \bibinfo {author} {\bibfnamefont {S.}~\bibnamefont {Wanka}},
  \bibinfo {author} {\bibfnamefont {J.}~\bibnamefont {Hagel}}, \bibinfo
  {author} {\bibfnamefont {M.}~\bibnamefont {Reibelt}}, \bibinfo {author}
  {\bibfnamefont {D.}~\bibnamefont {Schweitzer}},\ and\ \bibinfo {author}
  {\bibfnamefont {J.}~\bibnamefont {Schlueter}},\ }\href
  {https://doi.org/https://doi.org/10.1016/S0379-6779(02)00406-X} {\bibfield
  {journal} {\bibinfo  {journal} {Synth. Met.}\ }\textbf {\bibinfo {volume}
  {133-134}},\ \bibinfo {pages} {201 } (\bibinfo {year} {2003})}\BibitemShut
  {NoStop}%
\bibitem [{\citenamefont {Vorontsov}\ \emph {et~al.}(2005)\citenamefont
  {Vorontsov}, \citenamefont {Sauls},\ and\ \citenamefont
  {Graf}}]{Vorontsov2005}%
  \BibitemOpen
  \bibfield  {author} {\bibinfo {author} {\bibfnamefont {A.~B.}\ \bibnamefont
  {Vorontsov}}, \bibinfo {author} {\bibfnamefont {J.~A.}\ \bibnamefont
  {Sauls}},\ and\ \bibinfo {author} {\bibfnamefont {M.~J.}\ \bibnamefont
  {Graf}},\ }\href {https://doi.org/10.1103/PhysRevB.72.184501} {\bibfield
  {journal} {\bibinfo  {journal} {Phys. Rev. B}\ }\textbf {\bibinfo {volume}
  {72}},\ \bibinfo {pages} {184501} (\bibinfo {year} {2005})}\BibitemShut
  {NoStop}%
\bibitem [{\citenamefont {Abrikosov}(1957)}]{Abrikosov1957}%
  \BibitemOpen
  \bibfield  {author} {\bibinfo {author} {\bibfnamefont {A.}~\bibnamefont
  {Abrikosov}},\ }\href@noop {} {\bibfield  {journal} {\bibinfo  {journal}
  {Eksp, Zh.: Teor. Fiz.}\ }\textbf {\bibinfo {volume} {32}},\ \bibinfo {pages}
  {1442} (\bibinfo {year} {1957})},\ \bibinfo {note} {[Sov. Phys. JETP 5, 1174
  (1957)]}\BibitemShut {NoStop}%
\bibitem [{\citenamefont {Houzet}\ \emph {et~al.}(2002)\citenamefont {Houzet},
  \citenamefont {Buzdin}, \citenamefont {Bulaevskii},\ and\ \citenamefont
  {Maley}}]{Houzet2002}%
  \BibitemOpen
  \bibfield  {author} {\bibinfo {author} {\bibfnamefont {M.}~\bibnamefont
  {Houzet}}, \bibinfo {author} {\bibfnamefont {A.}~\bibnamefont {Buzdin}},
  \bibinfo {author} {\bibfnamefont {L.}~\bibnamefont {Bulaevskii}},\ and\
  \bibinfo {author} {\bibfnamefont {M.}~\bibnamefont {Maley}},\ }\href
  {https://doi.org/10.1103/PhysRevLett.88.227001} {\bibfield  {journal}
  {\bibinfo  {journal} {Phys. Rev. Lett.}\ }\textbf {\bibinfo {volume} {88}},\
  \bibinfo {pages} {227001} (\bibinfo {year} {2002})}\BibitemShut {NoStop}%
\bibitem [{\citenamefont {Rainer}\ and\ \citenamefont
  {Bergmann}(1974)}]{Rainer1974}%
  \BibitemOpen
  \bibfield  {author} {\bibinfo {author} {\bibfnamefont {D.}~\bibnamefont
  {Rainer}}\ and\ \bibinfo {author} {\bibfnamefont {G.}~\bibnamefont
  {Bergmann}},\ }\href {https://doi.org/10.1007/BF00658876} {\bibfield
  {journal} {\bibinfo  {journal} {J. Low Temp. Phys.}\ }\textbf {\bibinfo
  {volume} {14}},\ \bibinfo {pages} {501} (\bibinfo {year} {1974})}\BibitemShut
  {NoStop}%
\bibitem [{\citenamefont {Sasaki}\ \emph {et~al.}(2002)\citenamefont {Sasaki},
  \citenamefont {Fukuda}, \citenamefont {Nishizaki}, \citenamefont {Fujita},
  \citenamefont {Yoneyama}, \citenamefont {Kobayashi},\ and\ \citenamefont
  {Biberacher}}]{Sasaki2002}%
  \BibitemOpen
  \bibfield  {author} {\bibinfo {author} {\bibfnamefont {T.}~\bibnamefont
  {Sasaki}}, \bibinfo {author} {\bibfnamefont {T.}~\bibnamefont {Fukuda}},
  \bibinfo {author} {\bibfnamefont {T.}~\bibnamefont {Nishizaki}}, \bibinfo
  {author} {\bibfnamefont {T.}~\bibnamefont {Fujita}}, \bibinfo {author}
  {\bibfnamefont {N.}~\bibnamefont {Yoneyama}}, \bibinfo {author}
  {\bibfnamefont {N.}~\bibnamefont {Kobayashi}},\ and\ \bibinfo {author}
  {\bibfnamefont {W.}~\bibnamefont {Biberacher}},\ }\href
  {https://doi.org/10.1103/PhysRevB.66.224513} {\bibfield  {journal} {\bibinfo
  {journal} {Phys. Rev. B}\ }\textbf {\bibinfo {volume} {66}},\ \bibinfo
  {pages} {224513} (\bibinfo {year} {2002})}\BibitemShut {NoStop}%
\bibitem [{\citenamefont {Houzet}\ and\ \citenamefont
  {Mineev}(2006)}]{Houzet2006}%
  \BibitemOpen
  \bibfield  {author} {\bibinfo {author} {\bibfnamefont {M.}~\bibnamefont
  {Houzet}}\ and\ \bibinfo {author} {\bibfnamefont {V.~P.}\ \bibnamefont
  {Mineev}},\ }\href {https://doi.org/10.1103/PhysRevB.74.144522} {\bibfield
  {journal} {\bibinfo  {journal} {Phys. Rev. B}\ }\textbf {\bibinfo {volume}
  {74}},\ \bibinfo {pages} {144522} (\bibinfo {year} {2006})}\BibitemShut
  {NoStop}%
\bibitem [{\citenamefont {Mola}\ \emph {et~al.}(2001)\citenamefont {Mola},
  \citenamefont {Hill}, \citenamefont {Brooks},\ and\ \citenamefont
  {Qualls}}]{Mola2001}%
  \BibitemOpen
  \bibfield  {author} {\bibinfo {author} {\bibfnamefont {M.~M.}\ \bibnamefont
  {Mola}}, \bibinfo {author} {\bibfnamefont {S.}~\bibnamefont {Hill}}, \bibinfo
  {author} {\bibfnamefont {J.~S.}\ \bibnamefont {Brooks}},\ and\ \bibinfo
  {author} {\bibfnamefont {J.~S.}\ \bibnamefont {Qualls}},\ }\href
  {https://doi.org/10.1103/PhysRevLett.86.2130} {\bibfield  {journal} {\bibinfo
   {journal} {Phys. Rev. Lett.}\ }\textbf {\bibinfo {volume} {86}},\ \bibinfo
  {pages} {2130} (\bibinfo {year} {2001})}\BibitemShut {NoStop}%
\bibitem [{\citenamefont {Konoike}\ \emph {et~al.}(2009)\citenamefont
  {Konoike}, \citenamefont {Uchida}, \citenamefont {Osada}, \citenamefont
  {Yamaguchi}, \citenamefont {Nishimura}, \citenamefont {Terashima},
  \citenamefont {Uji},\ and\ \citenamefont {Yamada}}]{Konoike2009}%
  \BibitemOpen
  \bibfield  {author} {\bibinfo {author} {\bibfnamefont {T.}~\bibnamefont
  {Konoike}}, \bibinfo {author} {\bibfnamefont {K.}~\bibnamefont {Uchida}},
  \bibinfo {author} {\bibfnamefont {T.}~\bibnamefont {Osada}}, \bibinfo
  {author} {\bibfnamefont {T.}~\bibnamefont {Yamaguchi}}, \bibinfo {author}
  {\bibfnamefont {M.}~\bibnamefont {Nishimura}}, \bibinfo {author}
  {\bibfnamefont {T.}~\bibnamefont {Terashima}}, \bibinfo {author}
  {\bibfnamefont {S.}~\bibnamefont {Uji}},\ and\ \bibinfo {author}
  {\bibfnamefont {J.}~\bibnamefont {Yamada}},\ }\href
  {https://doi.org/10.1103/PhysRevB.79.054509} {\bibfield  {journal} {\bibinfo
  {journal} {Phys. Rev. B}\ }\textbf {\bibinfo {volume} {79}},\ \bibinfo
  {pages} {054509} (\bibinfo {year} {2009})}\BibitemShut {NoStop}%
\bibitem [{\citenamefont {Yin}\ \emph {et~al.}(2007)\citenamefont {Yin},
  \citenamefont {Nam}, \citenamefont {Analytis}, \citenamefont {Blundell},
  \citenamefont {Ardavan}, \citenamefont {Schlueter},\ and\ \citenamefont
  {Sasaki}}]{Yin2007}%
  \BibitemOpen
  \bibfield  {author} {\bibinfo {author} {\bibfnamefont {L.}~\bibnamefont
  {Yin}}, \bibinfo {author} {\bibfnamefont {M.-S.}\ \bibnamefont {Nam}},
  \bibinfo {author} {\bibfnamefont {J.~G.}\ \bibnamefont {Analytis}}, \bibinfo
  {author} {\bibfnamefont {S.~J.}\ \bibnamefont {Blundell}}, \bibinfo {author}
  {\bibfnamefont {A.}~\bibnamefont {Ardavan}}, \bibinfo {author} {\bibfnamefont
  {J.~A.}\ \bibnamefont {Schlueter}},\ and\ \bibinfo {author} {\bibfnamefont
  {T.}~\bibnamefont {Sasaki}},\ }\href
  {https://doi.org/10.1103/PhysRevB.76.014506} {\bibfield  {journal} {\bibinfo
  {journal} {Phys. Rev. B}\ }\textbf {\bibinfo {volume} {76}},\ \bibinfo
  {pages} {014506} (\bibinfo {year} {2007})}\BibitemShut {NoStop}%
\bibitem [{\citenamefont {Ashauer}\ \emph {et~al.}(1987)\citenamefont
  {Ashauer}, \citenamefont {Lee},\ and\ \citenamefont {Rammer}}]{Ashauer1987}%
  \BibitemOpen
  \bibfield  {author} {\bibinfo {author} {\bibfnamefont {B.}~\bibnamefont
  {Ashauer}}, \bibinfo {author} {\bibfnamefont {W.}~\bibnamefont {Lee}},\ and\
  \bibinfo {author} {\bibfnamefont {J.}~\bibnamefont {Rammer}},\ }\href
  {https://doi.org/10.1007/BF01303974} {\bibfield  {journal} {\bibinfo
  {journal} {Z. Phys. B}\ }\textbf {\bibinfo {volume} {67}},\ \bibinfo {pages}
  {147} (\bibinfo {year} {1987})}\BibitemShut {NoStop}%
\bibitem [{\citenamefont {Croitoru}\ and\ \citenamefont
  {Buzdin}(2017)}]{Croitoru2017}%
  \BibitemOpen
  \bibfield  {author} {\bibinfo {author} {\bibfnamefont {M.~D.}\ \bibnamefont
  {Croitoru}}\ and\ \bibinfo {author} {\bibfnamefont {A.~I.}\ \bibnamefont
  {Buzdin}},\ }\href@noop {} {\bibfield  {journal} {\bibinfo  {journal}
  {Condens. Matter}\ }\textbf {\bibinfo {volume} {2}} (\bibinfo {year}
  {2017})}\BibitemShut {NoStop}%
\bibitem [{\citenamefont {Shimahara}(1998)}]{Shimahara1998}%
  \BibitemOpen
  \bibfield  {author} {\bibinfo {author} {\bibfnamefont {H.}~\bibnamefont
  {Shimahara}},\ }\href {https://doi.org/10.1143/JPSJ.67.1872} {\bibfield
  {journal} {\bibinfo  {journal} {J. Phys. Soc. Jpn.}\ }\textbf {\bibinfo
  {volume} {67}},\ \bibinfo {pages} {1872} (\bibinfo {year}
  {1998})}\BibitemShut {NoStop}%
\bibitem [{\citenamefont {Zwicknagl}\ \emph {et~al.}(2017)\citenamefont
  {Zwicknagl}, \citenamefont {Jahns},\ and\ \citenamefont
  {Fulde}}]{Zwicknagl2017}%
  \BibitemOpen
  \bibfield  {author} {\bibinfo {author} {\bibfnamefont {G.}~\bibnamefont
  {Zwicknagl}}, \bibinfo {author} {\bibfnamefont {S.}~\bibnamefont {Jahns}},\
  and\ \bibinfo {author} {\bibfnamefont {P.}~\bibnamefont {Fulde}},\
  }\href@noop {} {\bibfield  {journal} {\bibinfo  {journal} {J. Phys. Soc.
  Jpn.}\ }\textbf {\bibinfo {volume} {86}},\ \bibinfo {pages} {083701}
  (\bibinfo {year} {2017})}\BibitemShut {NoStop}%
\bibitem [{\citenamefont {Brandt}(1995)}]{Brandt1995}%
  \BibitemOpen
  \bibfield  {author} {\bibinfo {author} {\bibfnamefont {E.~H.}\ \bibnamefont
  {Brandt}},\ }\href {https://doi.org/10.1088/0034-4885/58/11/003} {\bibfield
  {journal} {\bibinfo  {journal} {Rep. Prog. Phys.}\ }\textbf {\bibinfo
  {volume} {58}},\ \bibinfo {pages} {1465} (\bibinfo {year}
  {1995})}\BibitemShut {NoStop}%
\end{thebibliography}%


\begin{thebibliography}{24}%
\makeatletter
\providecommand \@ifxundefined [1]{%
 \@ifx{#1\undefined}
}%
\providecommand \@ifnum [1]{%
 \ifnum #1\expandafter \@firstoftwo
 \else \expandafter \@secondoftwo
 \fi
}%
\providecommand \@ifx [1]{%
 \ifx #1\expandafter \@firstoftwo
 \else \expandafter \@secondoftwo
 \fi
}%
\providecommand \natexlab [1]{#1}%
\providecommand \enquote  [1]{``#1''}%
\providecommand \bibnamefont  [1]{#1}%
\providecommand \bibfnamefont [1]{#1}%
\providecommand \citenamefont [1]{#1}%
\providecommand \href@noop [0]{\@secondoftwo}%
\providecommand \href [0]{\begingroup \@sanitize@url \@href}%
\providecommand \@href[1]{\@@startlink{#1}\@@href}%
\providecommand \@@href[1]{\endgroup#1\@@endlink}%
\providecommand \@sanitize@url [0]{\catcode `\\12\catcode `\$12\catcode
  `\&12\catcode `\#12\catcode `\^12\catcode `\_12\catcode `\%12\relax}%
\providecommand \@@startlink[1]{}%
\providecommand \@@endlink[0]{}%
\providecommand \url  [0]{\begingroup\@sanitize@url \@url }%
\providecommand \@url [1]{\endgroup\@href {#1}{\urlprefix }}%
\providecommand \urlprefix  [0]{URL }%
\providecommand \Eprint [0]{\href }%
\providecommand \doibase [0]{https://doi.org/}%
\providecommand \selectlanguage [0]{\@gobble}%
\providecommand \bibinfo  [0]{\@secondoftwo}%
\providecommand \bibfield  [0]{\@secondoftwo}%
\providecommand \translation [1]{[#1]}%
\providecommand \BibitemOpen [0]{}%
\providecommand \bibitemStop [0]{}%
\providecommand \bibitemNoStop [0]{.\EOS\space}%
\providecommand \EOS [0]{\spacefactor3000\relax}%
\providecommand \BibitemShut  [1]{\csname bibitem#1\endcsname}%
\let\auto@bib@innerbib\@empty
\bibitem [{\citenamefont {Wosnitza}(1999)}]{Wosnitza1999}%
  \BibitemOpen
  \bibfield  {author} {\bibinfo {author} {\bibfnamefont {J.}~\bibnamefont
  {Wosnitza}},\ }\href {https://doi.org/10.1023/A:1022592522487} {\bibfield
  {journal} {\bibinfo  {journal} {J. Low Temp. Phys.}\ }\textbf {\bibinfo
  {volume} {117}},\ \bibinfo {pages} {1701} (\bibinfo {year}
  {1999})}\BibitemShut {NoStop}%
\bibitem [{\citenamefont {Won}\ and\ \citenamefont {Maki}(1995)}]{Won1995}%
  \BibitemOpen
  \bibfield  {author} {\bibinfo {author} {\bibfnamefont {H.}~\bibnamefont
  {Won}}\ and\ \bibinfo {author} {\bibfnamefont {K.}~\bibnamefont {Maki}},\
  }\href {https://doi.org/10.1209/0295-5075/30/7/008} {\bibfield  {journal}
  {\bibinfo  {journal} {EPL}\ }\textbf {\bibinfo {volume} {30}},\ \bibinfo
  {pages} {421} (\bibinfo {year} {1995})}\BibitemShut {NoStop}%
\bibitem [{\citenamefont {Won}\ \emph {et~al.}(2004)\citenamefont {Won},
  \citenamefont {Haas},\ and\ \citenamefont {Maki}}]{Won2004}%
  \BibitemOpen
  \bibfield  {author} {\bibinfo {author} {\bibfnamefont {H.}~\bibnamefont
  {Won}}, \bibinfo {author} {\bibfnamefont {S.}~\bibnamefont {Haas}},\ and\
  \bibinfo {author} {\bibfnamefont {K.}~\bibnamefont {Maki}},\ }\href
  {https://doi.org/https://doi.org/10.1016/j.jmmm.2004.03.027} {\bibfield
  {journal} {\bibinfo  {journal} {J. Magn. Magn. Mater.}\ }\textbf {\bibinfo
  {volume} {272-276}},\ \bibinfo {pages} {191} (\bibinfo {year} {2004})},\
  \bibinfo {note} {proceedings of the International Conference on Magnetism
  (ICM 2003)}\BibitemShut {NoStop}%
\bibitem [{\citenamefont {Won}\ and\ \citenamefont {Maki}(2001)}]{Won2001}%
  \BibitemOpen
  \bibfield  {author} {\bibinfo {author} {\bibfnamefont {H.}~\bibnamefont
  {Won}}\ and\ \bibinfo {author} {\bibfnamefont {K.}~\bibnamefont {Maki}},\
  }\href {https://doi.org/10.1209/epl/i2001-00581-4} {\bibfield  {journal}
  {\bibinfo  {journal} {EPL}\ }\textbf {\bibinfo {volume} {56}},\ \bibinfo
  {pages} {729} (\bibinfo {year} {2001})}\BibitemShut {NoStop}%
\bibitem [{\citenamefont {Lebed}\ and\ \citenamefont
  {Sepper}(2020)}]{Lebed2020}%
  \BibitemOpen
  \bibfield  {author} {\bibinfo {author} {\bibfnamefont {A.~G.}\ \bibnamefont
  {Lebed}}\ and\ \bibinfo {author} {\bibfnamefont {O.}~\bibnamefont {Sepper}},\
  }\href {https://doi.org/10.1134/S0021364020040037} {\bibfield  {journal}
  {\bibinfo  {journal} {JETP Lett.}\ }\textbf {\bibinfo {volume} {111}},\
  \bibinfo {pages} {239} (\bibinfo {year} {2020})}\BibitemShut {NoStop}%
\bibitem [{\citenamefont {Lebed}(2021)}]{Lebed2021}%
  \BibitemOpen
  \bibfield  {author} {\bibinfo {author} {\bibfnamefont {A.~G.}\ \bibnamefont
  {Lebed}},\ }\href {https://doi.org/10.1134/S0021364021110011} {\bibfield
  {journal} {\bibinfo  {journal} {JETP Lett.}\ }\textbf {\bibinfo {volume}
  {113}},\ \bibinfo {pages} {701} (\bibinfo {year} {2021})}\BibitemShut
  {NoStop}%
\bibitem [{\citenamefont {Zwicknagl}\ and\ \citenamefont
  {Wosnitza}(2010)}]{Zwicknagl2010}%
  \BibitemOpen
  \bibfield  {author} {\bibinfo {author} {\bibfnamefont {G.}~\bibnamefont
  {Zwicknagl}}\ and\ \bibinfo {author} {\bibfnamefont {J.}~\bibnamefont
  {Wosnitza}},\ }\href {https://doi.org/10.1142/S0217979210056396} {\bibfield
  {journal} {\bibinfo  {journal} {Int. J. Mod. Phys. B}\ }\textbf {\bibinfo
  {volume} {24}},\ \bibinfo {pages} {3915} (\bibinfo {year}
  {2010})}\BibitemShut {NoStop}%
\bibitem [{\citenamefont {Shimahara}\ and\ \citenamefont
  {Rainer}(1997)}]{Shimahara1997}%
  \BibitemOpen
  \bibfield  {author} {\bibinfo {author} {\bibfnamefont {H.}~\bibnamefont
  {Shimahara}}\ and\ \bibinfo {author} {\bibfnamefont {D.}~\bibnamefont
  {Rainer}},\ }\href {https://doi.org/10.1143/JPSJ.66.3591} {\bibfield
  {journal} {\bibinfo  {journal} {J. Phys. Soc. Jpn.}\ }\textbf {\bibinfo
  {volume} {66}},\ \bibinfo {pages} {3591} (\bibinfo {year}
  {1997})}\BibitemShut {NoStop}%
\bibitem [{\citenamefont {Lortz}\ \emph {et~al.}(2007)\citenamefont {Lortz},
  \citenamefont {Wang}, \citenamefont {Demuer}, \citenamefont {B\"ottger},
  \citenamefont {Bergk}, \citenamefont {Zwicknagl}, \citenamefont {Nakazawa},\
  and\ \citenamefont {Wosnitza}}]{Lortz2007}%
  \BibitemOpen
  \bibfield  {author} {\bibinfo {author} {\bibfnamefont {R.}~\bibnamefont
  {Lortz}}, \bibinfo {author} {\bibfnamefont {Y.}~\bibnamefont {Wang}},
  \bibinfo {author} {\bibfnamefont {A.}~\bibnamefont {Demuer}}, \bibinfo
  {author} {\bibfnamefont {P.~H.~M.}\ \bibnamefont {B\"ottger}}, \bibinfo
  {author} {\bibfnamefont {B.}~\bibnamefont {Bergk}}, \bibinfo {author}
  {\bibfnamefont {G.}~\bibnamefont {Zwicknagl}}, \bibinfo {author}
  {\bibfnamefont {Y.}~\bibnamefont {Nakazawa}},\ and\ \bibinfo {author}
  {\bibfnamefont {J.}~\bibnamefont {Wosnitza}},\ }\href
  {https://doi.org/10.1103/PhysRevLett.99.187002} {\bibfield  {journal}
  {\bibinfo  {journal} {Phys. Rev. Lett.}\ }\textbf {\bibinfo {volume} {99}},\
  \bibinfo {pages} {187002} (\bibinfo {year} {2007})}\BibitemShut {NoStop}%
\bibitem [{\citenamefont {Rieck}\ and\ \citenamefont
  {Scharnberg}(1990)}]{Rieck1990}%
  \BibitemOpen
  \bibfield  {author} {\bibinfo {author} {\bibfnamefont {C.~T.}\ \bibnamefont
  {Rieck}}\ and\ \bibinfo {author} {\bibfnamefont {K.}~\bibnamefont
  {Scharnberg}},\ }\href@noop {} {\bibfield  {journal} {\bibinfo  {journal}
  {Physica B}\ }\textbf {\bibinfo {volume} {163}},\ \bibinfo {pages} {670}
  (\bibinfo {year} {1990})}\BibitemShut {NoStop}%
\bibitem [{\citenamefont {Mola}\ \emph {et~al.}(2001)\citenamefont {Mola},
  \citenamefont {Hill}, \citenamefont {Brooks},\ and\ \citenamefont
  {Qualls}}]{Mola2001}%
  \BibitemOpen
  \bibfield  {author} {\bibinfo {author} {\bibfnamefont {M.~M.}\ \bibnamefont
  {Mola}}, \bibinfo {author} {\bibfnamefont {S.}~\bibnamefont {Hill}}, \bibinfo
  {author} {\bibfnamefont {J.~S.}\ \bibnamefont {Brooks}},\ and\ \bibinfo
  {author} {\bibfnamefont {J.~S.}\ \bibnamefont {Qualls}},\ }\href
  {https://doi.org/10.1103/PhysRevLett.86.2130} {\bibfield  {journal} {\bibinfo
   {journal} {Phys. Rev. Lett.}\ }\textbf {\bibinfo {volume} {86}},\ \bibinfo
  {pages} {2130} (\bibinfo {year} {2001})}\BibitemShut {NoStop}%
\bibitem [{\citenamefont {Sasaki}\ \emph {et~al.}(2002)\citenamefont {Sasaki},
  \citenamefont {Fukuda}, \citenamefont {Nishizaki}, \citenamefont {Fujita},
  \citenamefont {Yoneyama}, \citenamefont {Kobayashi},\ and\ \citenamefont
  {Biberacher}}]{Sasaki2002}%
  \BibitemOpen
  \bibfield  {author} {\bibinfo {author} {\bibfnamefont {T.}~\bibnamefont
  {Sasaki}}, \bibinfo {author} {\bibfnamefont {T.}~\bibnamefont {Fukuda}},
  \bibinfo {author} {\bibfnamefont {T.}~\bibnamefont {Nishizaki}}, \bibinfo
  {author} {\bibfnamefont {T.}~\bibnamefont {Fujita}}, \bibinfo {author}
  {\bibfnamefont {N.}~\bibnamefont {Yoneyama}}, \bibinfo {author}
  {\bibfnamefont {N.}~\bibnamefont {Kobayashi}},\ and\ \bibinfo {author}
  {\bibfnamefont {W.}~\bibnamefont {Biberacher}},\ }\href
  {https://doi.org/10.1103/PhysRevB.66.224513} {\bibfield  {journal} {\bibinfo
  {journal} {Phys. Rev. B}\ }\textbf {\bibinfo {volume} {66}},\ \bibinfo
  {pages} {224513} (\bibinfo {year} {2002})}\BibitemShut {NoStop}%
\bibitem [{\citenamefont {Uji}\ \emph {et~al.}(2018{\natexlab{a}})\citenamefont
  {Uji}, \citenamefont {Fujii}, \citenamefont {Sugiura}, \citenamefont
  {Terashima}, \citenamefont {Isono},\ and\ \citenamefont {Yamada}}]{Uji2018a}%
  \BibitemOpen
  \bibfield  {author} {\bibinfo {author} {\bibfnamefont {S.}~\bibnamefont
  {Uji}}, \bibinfo {author} {\bibfnamefont {Y.}~\bibnamefont {Fujii}}, \bibinfo
  {author} {\bibfnamefont {S.}~\bibnamefont {Sugiura}}, \bibinfo {author}
  {\bibfnamefont {T.}~\bibnamefont {Terashima}}, \bibinfo {author}
  {\bibfnamefont {T.}~\bibnamefont {Isono}},\ and\ \bibinfo {author}
  {\bibfnamefont {J.}~\bibnamefont {Yamada}},\ }\href
  {https://doi.org/10.1103/PhysRevB.97.024505} {\bibfield  {journal} {\bibinfo
  {journal} {Phys. Rev. B}\ }\textbf {\bibinfo {volume} {97}},\ \bibinfo
  {pages} {024505} (\bibinfo {year} {2018}{\natexlab{a}})}\BibitemShut
  {NoStop}%
\bibitem [{\citenamefont {Bateman}\ and\ \citenamefont
  {Erdélyi}(1955)}]{Bateman1955}%
  \BibitemOpen
  \bibfield  {author} {\bibinfo {author} {\bibfnamefont {H.}~\bibnamefont
  {Bateman}}\ and\ \bibinfo {author} {\bibfnamefont {A.}~\bibnamefont
  {Erdélyi}},\ }\href {https://cds.cern.ch/record/100233} {\emph {\bibinfo
  {title} {{Higher transcendental functions}}}},\ California Institute of
  technology. Bateman Manuscript project\ (\bibinfo  {publisher}
  {McGraw-Hill},\ \bibinfo {address} {New York, NY},\ \bibinfo {year}
  {1955})\BibitemShut {NoStop}%
\bibitem [{\citenamefont {Zwicknagl}\ \emph {et~al.}(2017)\citenamefont
  {Zwicknagl}, \citenamefont {Jahns},\ and\ \citenamefont
  {Fulde}}]{Zwicknagl2017}%
  \BibitemOpen
  \bibfield  {author} {\bibinfo {author} {\bibfnamefont {G.}~\bibnamefont
  {Zwicknagl}}, \bibinfo {author} {\bibfnamefont {S.}~\bibnamefont {Jahns}},\
  and\ \bibinfo {author} {\bibfnamefont {P.}~\bibnamefont {Fulde}},\
  }\href@noop {} {\bibfield  {journal} {\bibinfo  {journal} {J. Phys. Soc.
  Jpn.}\ }\textbf {\bibinfo {volume} {86}},\ \bibinfo {pages} {083701}
  (\bibinfo {year} {2017})}\BibitemShut {NoStop}%
\bibitem [{\citenamefont {Agosta}\ \emph {et~al.}(2012)\citenamefont {Agosta},
  \citenamefont {Jin}, \citenamefont {Coniglio}, \citenamefont {Smith},
  \citenamefont {Cho}, \citenamefont {Stroe}, \citenamefont {Martin},
  \citenamefont {Tozer}, \citenamefont {Murphy}, \citenamefont {Palm},
  \citenamefont {Schlueter},\ and\ \citenamefont {Kurmoo}}]{Agosta2012}%
  \BibitemOpen
  \bibfield  {author} {\bibinfo {author} {\bibfnamefont {C.~C.}\ \bibnamefont
  {Agosta}}, \bibinfo {author} {\bibfnamefont {J.}~\bibnamefont {Jin}},
  \bibinfo {author} {\bibfnamefont {W.~A.}\ \bibnamefont {Coniglio}}, \bibinfo
  {author} {\bibfnamefont {B.~E.}\ \bibnamefont {Smith}}, \bibinfo {author}
  {\bibfnamefont {K.}~\bibnamefont {Cho}}, \bibinfo {author} {\bibfnamefont
  {I.}~\bibnamefont {Stroe}}, \bibinfo {author} {\bibfnamefont
  {C.}~\bibnamefont {Martin}}, \bibinfo {author} {\bibfnamefont {S.~W.}\
  \bibnamefont {Tozer}}, \bibinfo {author} {\bibfnamefont {T.~P.}\ \bibnamefont
  {Murphy}}, \bibinfo {author} {\bibfnamefont {E.~C.}\ \bibnamefont {Palm}},
  \bibinfo {author} {\bibfnamefont {J.~A.}\ \bibnamefont {Schlueter}},\ and\
  \bibinfo {author} {\bibfnamefont {M.}~\bibnamefont {Kurmoo}},\ }\href
  {https://doi.org/10.1103/PhysRevB.85.214514} {\bibfield  {journal} {\bibinfo
  {journal} {Phys. Rev. B}\ }\textbf {\bibinfo {volume} {85}},\ \bibinfo
  {pages} {214514} (\bibinfo {year} {2012})}\BibitemShut {NoStop}%
\bibitem [{\citenamefont {Bergk}\ \emph {et~al.}(2011)\citenamefont {Bergk},
  \citenamefont {Demuer}, \citenamefont {Sheikin}, \citenamefont {Wang},
  \citenamefont {Wosnitza}, \citenamefont {Nakazawa},\ and\ \citenamefont
  {Lortz}}]{Bergk2011}%
  \BibitemOpen
  \bibfield  {author} {\bibinfo {author} {\bibfnamefont {B.}~\bibnamefont
  {Bergk}}, \bibinfo {author} {\bibfnamefont {A.}~\bibnamefont {Demuer}},
  \bibinfo {author} {\bibfnamefont {I.}~\bibnamefont {Sheikin}}, \bibinfo
  {author} {\bibfnamefont {Y.}~\bibnamefont {Wang}}, \bibinfo {author}
  {\bibfnamefont {J.}~\bibnamefont {Wosnitza}}, \bibinfo {author}
  {\bibfnamefont {Y.}~\bibnamefont {Nakazawa}},\ and\ \bibinfo {author}
  {\bibfnamefont {R.}~\bibnamefont {Lortz}},\ }\href
  {https://doi.org/10.1103/PhysRevB.83.064506} {\bibfield  {journal} {\bibinfo
  {journal} {Phys. Rev. B}\ }\textbf {\bibinfo {volume} {83}},\ \bibinfo
  {pages} {064506} (\bibinfo {year} {2011})}\BibitemShut {NoStop}%
\bibitem [{\citenamefont {Bulaevskii}\ \emph {et~al.}(2003)\citenamefont
  {Bulaevskii}, \citenamefont {Buzdin},\ and\ \citenamefont
  {Maley}}]{Bulaevskii2003}%
  \BibitemOpen
  \bibfield  {author} {\bibinfo {author} {\bibfnamefont {L.}~\bibnamefont
  {Bulaevskii}}, \bibinfo {author} {\bibfnamefont {A.}~\bibnamefont {Buzdin}},\
  and\ \bibinfo {author} {\bibfnamefont {M.}~\bibnamefont {Maley}},\ }\href
  {https://doi.org/10.1103/PhysRevLett.90.067003} {\bibfield  {journal}
  {\bibinfo  {journal} {Phys. Rev. Lett.}\ }\textbf {\bibinfo {volume} {90}},\
  \bibinfo {pages} {067003} (\bibinfo {year} {2003})}\BibitemShut {NoStop}%
\bibitem [{\citenamefont {Uji}\ \emph {et~al.}(2018{\natexlab{b}})\citenamefont
  {Uji}, \citenamefont {Iida}, \citenamefont {Sugiura}, \citenamefont {Isono},
  \citenamefont {Sugii}, \citenamefont {Kikugawa}, \citenamefont {Terashima},
  \citenamefont {Yasuzuka}, \citenamefont {Akutsu}, \citenamefont {Nakazawa},
  \citenamefont {Graf},\ and\ \citenamefont {Day}}]{Uji2018}%
  \BibitemOpen
  \bibfield  {author} {\bibinfo {author} {\bibfnamefont {S.}~\bibnamefont
  {Uji}}, \bibinfo {author} {\bibfnamefont {Y.}~\bibnamefont {Iida}}, \bibinfo
  {author} {\bibfnamefont {S.}~\bibnamefont {Sugiura}}, \bibinfo {author}
  {\bibfnamefont {T.}~\bibnamefont {Isono}}, \bibinfo {author} {\bibfnamefont
  {K.}~\bibnamefont {Sugii}}, \bibinfo {author} {\bibfnamefont
  {N.}~\bibnamefont {Kikugawa}}, \bibinfo {author} {\bibfnamefont
  {T.}~\bibnamefont {Terashima}}, \bibinfo {author} {\bibfnamefont
  {S.}~\bibnamefont {Yasuzuka}}, \bibinfo {author} {\bibfnamefont
  {H.}~\bibnamefont {Akutsu}}, \bibinfo {author} {\bibfnamefont
  {Y.}~\bibnamefont {Nakazawa}}, \bibinfo {author} {\bibfnamefont
  {D.}~\bibnamefont {Graf}},\ and\ \bibinfo {author} {\bibfnamefont
  {P.}~\bibnamefont {Day}},\ }\href
  {https://doi.org/10.1103/PhysRevB.97.144505} {\bibfield  {journal} {\bibinfo
  {journal} {Phys. Rev. B}\ }\textbf {\bibinfo {volume} {97}},\ \bibinfo
  {pages} {144505} (\bibinfo {year} {2018}{\natexlab{b}})}\BibitemShut
  {NoStop}%
\bibitem [{\citenamefont {Ashauer}\ \emph {et~al.}(1987)\citenamefont
  {Ashauer}, \citenamefont {Lee},\ and\ \citenamefont {Rammer}}]{Ashauer1987}%
  \BibitemOpen
  \bibfield  {author} {\bibinfo {author} {\bibfnamefont {B.}~\bibnamefont
  {Ashauer}}, \bibinfo {author} {\bibfnamefont {W.}~\bibnamefont {Lee}},\ and\
  \bibinfo {author} {\bibfnamefont {J.}~\bibnamefont {Rammer}},\ }\href
  {https://doi.org/10.1007/BF01303974} {\bibfield  {journal} {\bibinfo
  {journal} {Z. Phys. B}\ }\textbf {\bibinfo {volume} {67}},\ \bibinfo {pages}
  {147} (\bibinfo {year} {1987})}\BibitemShut {NoStop}%
\bibitem [{\citenamefont {Lebed}(2018)}]{Lebed2018}%
  \BibitemOpen
  \bibfield  {author} {\bibinfo {author} {\bibfnamefont {A.~G.}\ \bibnamefont
  {Lebed}},\ }\href {https://doi.org/10.1103/PhysRevB.97.144504} {\bibfield
  {journal} {\bibinfo  {journal} {Phys. Rev. B}\ }\textbf {\bibinfo {volume}
  {97}},\ \bibinfo {pages} {144504} (\bibinfo {year} {2018})}\BibitemShut
  {NoStop}%
\bibitem [{\citenamefont {Shimahara}(1998)}]{Shimahara1998}%
  \BibitemOpen
  \bibfield  {author} {\bibinfo {author} {\bibfnamefont {H.}~\bibnamefont
  {Shimahara}},\ }\href {https://doi.org/10.1143/JPSJ.67.1872} {\bibfield
  {journal} {\bibinfo  {journal} {J. Phys. Soc. Jpn.}\ }\textbf {\bibinfo
  {volume} {67}},\ \bibinfo {pages} {1872} (\bibinfo {year}
  {1998})}\BibitemShut {NoStop}%
\bibitem [{\citenamefont {Wang}\ \emph {et~al.}(2001)\citenamefont {Wang},
  \citenamefont {Plackowski},\ and\ \citenamefont {Junod}}]{Wang2001}%
  \BibitemOpen
  \bibfield  {author} {\bibinfo {author} {\bibfnamefont {Y.}~\bibnamefont
  {Wang}}, \bibinfo {author} {\bibfnamefont {T.}~\bibnamefont {Plackowski}},\
  and\ \bibinfo {author} {\bibfnamefont {A.}~\bibnamefont {Junod}},\ }\href
  {https://doi.org/https://doi.org/10.1016/S0921-4534(01)00617-7} {\bibfield
  {journal} {\bibinfo  {journal} {Physica C}\ }\textbf {\bibinfo {volume}
  {355}},\ \bibinfo {pages} {179 } (\bibinfo {year} {2001})}\BibitemShut
  {NoStop}%
\bibitem [{\citenamefont {Beyer}\ \emph {et~al.}(2012)\citenamefont {Beyer},
  \citenamefont {Bergk}, \citenamefont {Yasin}, \citenamefont {Schlueter},\
  and\ \citenamefont {Wosnitza}}]{Beyer2012}%
  \BibitemOpen
  \bibfield  {author} {\bibinfo {author} {\bibfnamefont {R.}~\bibnamefont
  {Beyer}}, \bibinfo {author} {\bibfnamefont {B.}~\bibnamefont {Bergk}},
  \bibinfo {author} {\bibfnamefont {S.}~\bibnamefont {Yasin}}, \bibinfo
  {author} {\bibfnamefont {J.~A.}\ \bibnamefont {Schlueter}},\ and\ \bibinfo
  {author} {\bibfnamefont {J.}~\bibnamefont {Wosnitza}},\ }\href
  {https://doi.org/10.1103/PhysRevLett.109.027003} {\bibfield  {journal}
  {\bibinfo  {journal} {Phys. Rev. Lett.}\ }\textbf {\bibinfo {volume} {109}},\
  \bibinfo {pages} {027003} (\bibinfo {year} {2012})}\BibitemShut {NoStop}%
\end{thebibliography}%

\end{document}



\title{Orbital-Induced Crossover of the Fulde-Ferrell-Larkin-Ovchinnikov Phase into Abrikosov-like States\\-Supplementary Information-}


\author{Tommy Kotte}
\email[]{t.kotte@hzdr.de}
\affiliation{Hochfeld-Magnetlabor Dresden (HLD-EMFL) and W\"urzburg-Dresden Cluster of Excellence ct.qmat, Helmholtz-Zentrum Dresden-Rossendorf, 01328 Dresden, Germany}
\author{Hannes Kühne}
\affiliation{Hochfeld-Magnetlabor Dresden (HLD-EMFL) and W\"urzburg-Dresden Cluster of Excellence ct.qmat, Helmholtz-Zentrum Dresden-Rossendorf, 01328 Dresden, Germany}
\author{John A Schlueter}
\affiliation{Materials Science Division, Argonne National Laboratory, Argonne, IL 60439, USA}
\affiliation{Division of Materials Research, National Science Foundation, Alexandria, VA 22314, USA}
\author{Gertrud Zwicknagl}
\affiliation{Institute for Mathematical Physics, Technische Universität Braunschweig, 38106 Braunschweig, Germany}
\author{J. Wosnitza}
\affiliation{Hochfeld-Magnetlabor Dresden (HLD-EMFL) and W\"urzburg-Dresden Cluster of Excellence ct.qmat, Helmholtz-Zentrum Dresden-Rossendorf, 01328 Dresden, Germany}
\affiliation{Institut f\"ur Festk\"orper- und Materialphysik, TU Dresden, 01062 Dresden, Germany}

\date{\today}

\begin{abstract}
In these Supplementary Information, we provide details on the calculation of the $H$-$T$ phase diagram for $\kappa$-(ET)$_2$Cu(NCS)$_2$ as function of the Landau index $n$ and the angle $\alpha$ of the applied field. In particular, we consider the possible impact of the pairing symmetry on the Abrikosov-like states and discuss additional effects which might alter the superconducting phase diagram further. Moreover, the experimental setup for the specific-heat measurements is introduced. Finally, we present additional experimental data, recorded at \SI{21.7}{\tesla} for negative and positive out-of-plane tilt angles as well as at \SI{19.5}{\tesla}.   
\end{abstract}


\maketitle
\section{Influence of the pairing mechanism}
The paring mechanism in the quasi-two-dimensional (Q2D) organic superconductors is still controversially discussed (see e.g. \cite{Wosnitza1999}). Our study does not aim to answer the controversy, whether the pairing mechanism in $\kappa$-(ET)$_2$Cu(NCS)$_2$ has $s$ or $d$-wave character. However, in these Supplementary Information, we like to point out the influence of the order-parameter symmetry on the high-field superconducting phase diagram and its impact on the higher-order Landau states which are discussed in our manuscript. 

In general, the SC order parameter shares the properties of a two-Fermion wave function: it depends on (i) the center-of-mass variable $\vec{R}=\frac{1}{2}\left(\vec{r}_1+\vec{r}_2\right)$ and (ii) the relative variable $\vec{r}=\vec{r}_1-\vec{r}_2$ of the particles at positions $\vec{r}_1$ and $\vec{r}_2$ that form the Cooper pairs. The length scale for variations with the center-off-mass variable of the order parameter is the coherence length which is much larger than the Fermi wavelength. Accordingly, the spatial modulation of the Abrikosov state and the FFLO phase involve the center-of-mass variation and one anticipates only a minor influence of the order-parameter symmetry on these states and the corresponding phase diagram. However, the classification of the order parameter according to its symmetry properties refers to variations with variable $\vec{r}$. The dependence on the relative variable is reflected in the variation with the angle on the Fermi surface. As a consequence, the modulation vector characterizing the FFLO state can have preferred directions relative to the nodal lines of a $d_{x^2-y^2}$-order parameter, which it has not for isotropic $s$-wave pairing symmetry.

The upper critical field for magnetic fields aligned parallel to the conducting layers has been studied in literature for both $s$-wave as well as for $d$-wave order \cite{Won1995,Won2004,Won2001,Lebed2020,Lebed2021,Zwicknagl2010}. In the case of $d_{x^2-y^2}$-symmetry, it is important to note that the relative orientation of the modulation vector $\vec{q}$ to the nodal lines of the order parameter changes at low temperatures (top left panel of Fig. \ref{Fig_S5}). While this crossing occurs outside the temperature range considered in the present experiment ($T\approx0.06T_c$), it becomes of importance when discussing the SC transition in tilted magnetic fields. 

Following these considerations and the comprehensive discussions given by Shimahara and Rainer \cite{Shimahara1997}, we calculated the critical fields for different tilt angles $\alpha$ assuming both $s$-wave as well as $d$-wave order-parameter symmetry.
  
\subsection*{Linearized self-consistency equation for $H_{c2}$ in the $s$-wave case}
The starting point is Eq. (2.9) in Ref. \cite{Shimahara1997}
\begin{widetext}
	\begin{align}
	-\log\left(\frac{T}{T_c}\right) &= \pi k_B T\int_0^\infty \frac{dt}{\sinh \left(\pi k_B T t\right)}\nonumber\\
	&\left\lbrace 1-\cos \left(ht\right)e^{\left[-\frac{1}{16}t^2\left(\hbar v_F\right)^2\kappa_\perp\right]}\sum_{k=0}^{n}\frac{\left(-1\right)^kn!}{\left(k!\right)^2\left(n-k\right)!}\left[\frac{1}{8}t^2\left(\hbar v_F\right)^2\kappa_\perp\right]^k\right\rbrace\mathrm{ ,}\label{Eq_Shimahara}
	\end{align}
\end{widetext}
in which we inserted the Boltzmann constant $k_B$ and the reduced Planck constant $\hbar$ that were set to unity in Ref. \cite{Shimahara1997}. In this equation, 
\begin{equation}
h = \frac{1}{2}g\mu_BH
\end{equation}
is the Zeeman energy in a magnetic field $H$ and 
\begin{equation}
\kappa_\perp = \frac{2e}{\hbar c}H\sin \alpha = \frac{2\pi}{\Phi_0}H\sin \alpha\mathrm{ ,}
\end{equation}
with the magnetic flux quantum $\Phi_0=2\pi\hbar c/2e$.
The $k$ sum in Eq. (\ref{Eq_Shimahara}) is Kummer's confluent hypergeometric function $_1F_1(-n;1;\frac{1}{8}t^2\left(\hbar v_F\right)^2\kappa_\perp)$ or the Laguerre polynomial $L_n(\frac{1}{8}t^2\left(\hbar v_F\right)^2\kappa_\perp)$. This observation greatly simplifies the calculation and allows us to systematically study the limit $n\rightarrow \infty$, where we expect to recover the usual FFLO solution (see Ref. \cite{Shimahara1997}).

By replacing the integration variable
\begin{equation}
\pi k_BTt\rightarrow u \mathrm{ ,}\nonumber
\end{equation}
we obtain for the expression on the right-hand side of Eq. (\ref{Eq_Shimahara})
\begin{widetext}
	\begin{equation}
\int_0^\infty\frac{du}{\sinh u}\left\lbrace 1-\cos\left(\frac{h}{\pi k_BT_c}\frac{T_c}{T}u\right) e^{\left[-\frac{1}{16}u^2\left(\frac{\hbar v_F}{\pi k_B T_c}\right)^2\left(\frac{T_c}{T}\right)^2\kappa_\perp\right]} L_n\left(\frac{1}{8}u^2\left(\frac{\hbar v_F}{\pi k_BT_c}\right)^2\left(\frac{T_c}{T}\right)^2\kappa_\perp\right)\right\rbrace\mathrm{ .}\label{Eq_selfconst}
	\end{equation}
\end{widetext}
Here, the properties of the material enter through the dimensionless ratios 
\begin{align}
\frac{H}{\bar{H}_Z}&=\frac{h}{\pi k_B T_c}\nonumber\\
\frac{H}{\bar{H}_{orb}} &= \left(\frac{\hbar v_F}{\pi k_B T_c}\right)^2\frac{2\pi}{\Phi_0}H\mathrm{ ,}\nonumber
\end{align}
which, assuming $g\approx2$, define characteristic fields
\begin{align}
\bar{H}_Z &= \frac{\pi k_B T_c}{\mu_B}\\
\bar{H}_{orb} &= \frac{\Phi_0}{2\pi}\left(\frac{\pi k_B T_c}{\hbar v_F}\right)^2\mathrm{ .}\label{Eq_Horb}
\end{align}
We evaluate $\bar{H}_Z$ directly and estimate $\bar{H}_{orb}$ from the initial slope of the upper cricital field in the vicinity of the superconducting transition temperature $T\rightarrow T_c$.

\subsubsection*{Initial slope of the upper critical field}
We start from the linearized self-consistency equation Eq. (\ref{Eq_Shimahara}) for magnetic field perpendicular to the layers and expand it to leading order in $H$ and $T-T_c$ in the vicinity of $T_c$. Using $H_{c2}=0$ at $T_c$ and recognizing the Riemann $\zeta$-function, we obtain
\begin{equation}
-\frac{1}{T_c} = \int_0^\infty \frac{du}{\sinh u}\left\lbrace\frac{1}{16}u^2\left(\frac{\hbar v_F}{\pi k_B T_c}\right)^2 \frac{2\pi}{\Phi_0}\left(\frac{dH_{c2}^{orb}}{dT}\right)_{T=T_c}\right\rbrace=\frac{1}{16}\frac{7}{2}\zeta\left(3\right)\frac{1}{\bar{H}_{orb}}\left(\frac{dH_{c2}^{orb}}{dT}\right)_{T=T_{c}}\nonumber
\end{equation}
yielding
\begin{equation}
\bar{H}_{orb} = \left(\frac{\pi k_B T_c}{\hbar v_F}\right)^2\frac{\Phi_0}{2\pi}=\frac{1}{16}\frac{7}{2}\zeta\left(3\right)\left[-T_c\left(\frac{dH^{orb}_{c2}}{dT}\right)_{T=T_c}\right]\mathrm{ .}
\end{equation}

\begin{figure}[bt]
	\includegraphics{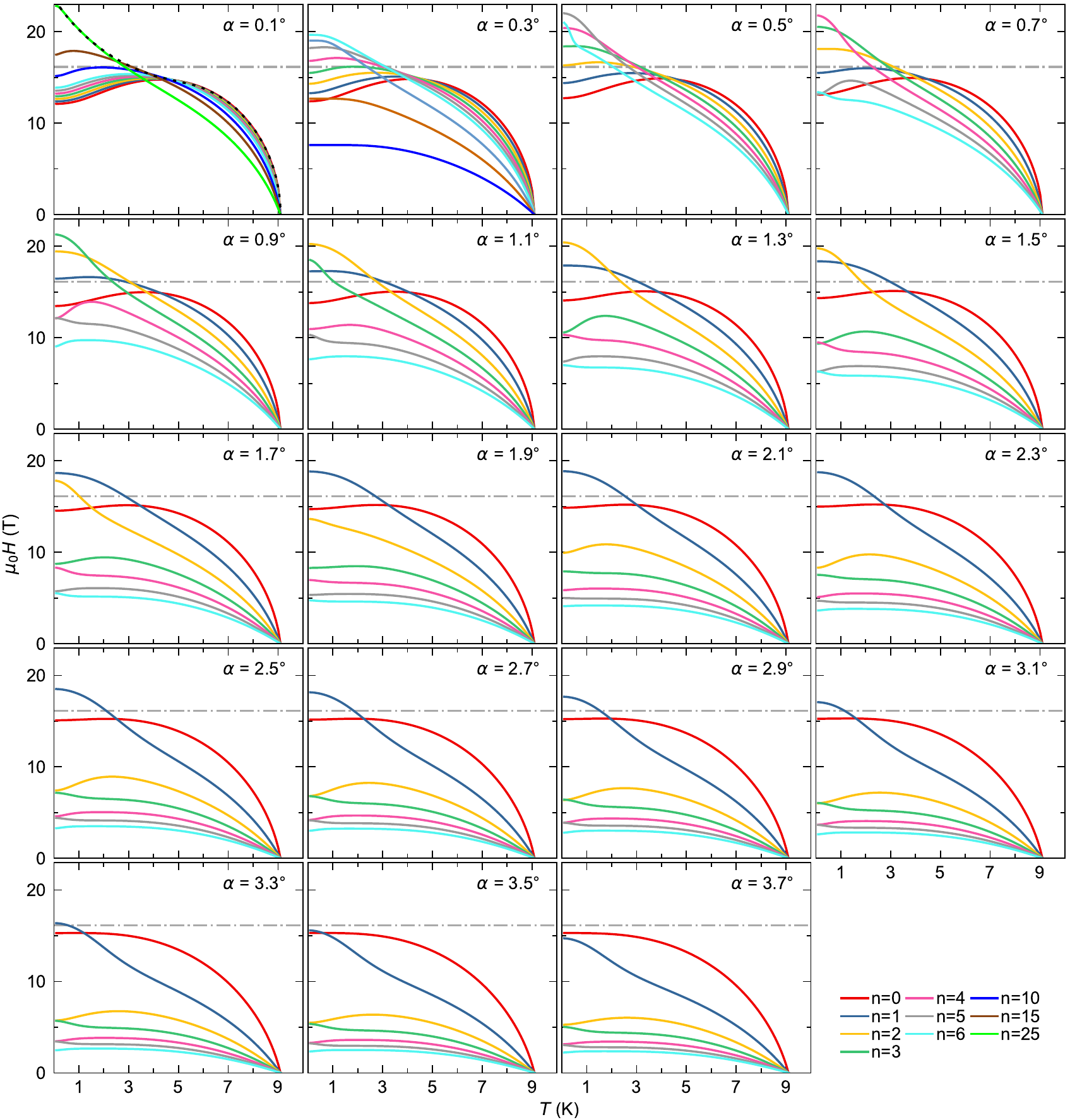}
	\caption{\label{Fig_S4} Calculated $H$-$T$ phase diagrams of $\kappa$-(ET)$_2$Cu(NCS)$_2$ for the lowest Landau levels and their variation with angle $\alpha$. Isotropic $s$-wave pairing is assumed. The ground state at a temperature $T$ is given by the Landau level with the highest critical field. The gray dashed-dotted line corresponds to the field of \SI{21.3}{\tesla} studied in this paper, if scaled by \SI{1.32}{} according to the strong-coupling enhancement factor of the thermodynamic critical field. The FFLO solution is plotted as black dashed line. For details see main text.}
\end{figure}

The $H$-$T$ phase diagrams presented in Fig. \ref{Fig_S4} were calculated with the value \mbox{$\left.\frac{dH^{orb}_{c2}}{dT}\right|_{T=T_c}=-\SI{0.66}{\tesla\per\kelvin}$}, estimated from the specific-heat data shown in Ref. \cite{Lortz2007}. Considering the clean limit of a quasi-two-dimensional (Q2D) superconductor with cylindrical Fermi-surface in a magnetic field applied perpendicular to the SC layers, this corresponds to an orbital critical field of \cite{Rieck1990}
\begin{equation}
H_{c2}^{orb}(T=0) = -0.59\cdot T_c\left.\frac{dH^{orb}_{c2}}{dT}\right|_{T=T_c} = \SI{3.5}{\tesla}\mathrm{ .}
\end{equation}
$H_{c2}^{orb}$ is of the order of previously reported values \cite{Lortz2007,Mola2001,Sasaki2002,Uji2018a} ranging from \SI{3}{} to \SI{7}{T}. A slight variation in $H_{c2}^{orb}$ does, however, not qualitatively change the $H$-$T$ phase diagram and our claims, but will only slightly alter the angular stability range of the high-field SC phase.

\subsubsection*{Recovery of the FFLO state}
In the following, we present a direct mathematical proof for the crossover from high-$n$ vortex states for small angles $\alpha$ into the usual FFLO phase as predicted by Ref. \cite{Shimahara1997}. The derivation is based on the leading term of the asymptotic expansion of the Laguerre polynomials \cite{Bateman1955} which converges uniformly in any compact interval of $\left(0,\infty\right)$
\begin{equation}
e^{-\frac{x}{2}}L_{n}\left(x\right)\to J_{0}\left(2\sqrt{\left(n+\frac{1}{2}\right)x}\right)\mathrm{ .}
\end{equation} 

With $x=\frac{u^2}{8}\left(\frac{\hbar v_F}{\pi k_B T}\right)^2\frac{2\pi}{\Phi_0}H_\perp$, we find for the right-hand side of the linearized self-consistency equation in the limit $n\rightarrow\infty$
\begin{equation}
\int_0^\infty\frac{du}{\sinh u}\left\lbrace 1-  \cos\left(\frac{h}{\pi k_B T}u\right)J_0\left(u\frac{1}{2}\frac{\hbar v_F}{\pi k_BT}\sqrt{2n\frac{2\pi}{\Phi_0}H_\perp}\right)\right\rbrace\mathrm{ .}
\end{equation}
This corresponds to the linearized self-consistency equation for an FFLO state with modulation vector $q=\sqrt{2n\frac{2\pi}{\Phi_0}H_\perp}$, which can be seen as follows: 
The upper critical field for the transition into an FFLO state with modulation vector of length $q$ is determined by the self-consistency equation
\begin{align}
-\log\frac{T}{T_c}=&\int_0^\infty\frac{du}{\sinh u}\left\lbrace 1-\frac{1}{2\pi}\int_0^{2\pi}d\phi \cos\left(\frac{h+\frac{1}{2}\hbar v_F q \cos\phi}{\pi k_B T}u\right)\right\rbrace\nonumber\\
=&\int_0^\infty\frac{du}{\sinh u}\left\lbrace 1-\cos\left(\frac{h}{\pi k_B T}u\right)J_0\left(u\frac{1}{2}\frac{\hbar v_F q}{\pi k_B T}\right) \right\rbrace\mathrm{ .}\label{Eq_FFLOap}
\end{align}
The approximate linearized self-consistency equation, Eq. (\ref{Eq_FFLOap}), is used to estimate the upper critical field for large quantum numbers $n$ and small angles $\alpha$.
\newpage
\subsection*{Linearized self-consistency equation for $H_{c2}$ in the $d$-wave case}
As a consequence of the anisotropic pairing interaction underlying $d_{x^2-y^2}$pairing, the superconducting states forming at a second-order transition have to be superpositions of harmonic oscillator states $\phi_n\left(\vec{r}\right)$ \cite{Shimahara1997}
\begin{equation}
\Delta_d\left(\vec{r}\right)=\sum^\infty_{n=0}\Delta_d^n\phi_n\left(\vec{r}\right)\mathrm{ .}\label{lin_self_d}
\end{equation}
The coefficients $\Delta_d^n$ are determined by the linearized self-consistency equation
\begin{equation}
-\log\frac{T}{T_c}\Delta_d^n = \sum^\infty_{n'=0}D_{nn'}\Delta_d^{n'}\mathrm{ ,}\label{Eigenwert}
\end{equation}
in which the matrix elements $D_{nn'}$ are given by
\begin{equation}
D_{nn'}=D_{nn}\delta_{nn'}+\left(\delta_{n',n-4}+\delta_{n',n+4}\right)D_{n,n'}^{(4)}\mathrm{ .}
\end{equation}
In contrast to the case of isotropic superconductors, the index $n$ is hence not a ``good" quantum number and the states will be superpositions reflecting the symmetry of the anisotropic order parameter. The change in the direction of the $\vec{q}$-vector relative to the nodal direction, as predicted for the pure FFLO state, could be reflected in a change of this superposition.

The diagonal terms are identical to the $n$-dependent factor appearing in the linearized self-consistency equation of $s$-wave superconductors [Eq. (\ref{Eq_selfconst})]
\begin{equation}
D_{nn}=k_B T\int_0^\infty \frac{dt}{\sinh \left(\pi k_B T t\right)}\left\lbrace 1-\cos \left(ht\right)e^{\left[-\frac{1}{16}t^2\left(\hbar v_F\right)^2\kappa_\perp\right]}L_n\left(\frac{1}{8}t^2\left(\hbar v_F\right)^2\kappa_\perp\right)\right\rbrace\mathrm{ .}
\end{equation}
The off-diagonal matrix elements can be rewritten as
\begin{align}
D_{n,n+4}^{(4)}&=\left(D_{n+4,n}^{(4)}\right)^*\nonumber\\
&=-e^{i4\alpha}\frac{\pi k_B T}{2}\int^{\infty}_0\frac{dt}{\sinh\left(\pi k_BTt\right)}e^{\left[-\frac{1}{16}t^2\left(\hbar v_F \right)^2\kappa_\perp\right]}\left(\frac{1}{8}t^2\left(\hbar v_F\right)^2\kappa_\perp\right)^2L_n^{(4)}\left(\frac{1}{8}t^2\left(\hbar v_F\right)^2\kappa_\perp\right)\label{off_d}\mathrm{ ,}
\end{align}
with the associated Laguerre-polynomials $L_n^{(4)}\left(x\right)$.
The upper critical field is given by the order parameter Eq. (\ref{lin_self_d}) where the coefficients $\Delta_d^n$ are the components of the eigenvector of Eq. (\ref{Eigenwert}) with the lowest eigenvalue.
As can be seen from Eq. (\ref{off_d}), the eigenvectors will contain contributions from $n_0$, $n_0+4$, $n_0+8, ...$, with $n_0=0,1,2,3$.

\begin{figure}[bt]
	\includegraphics{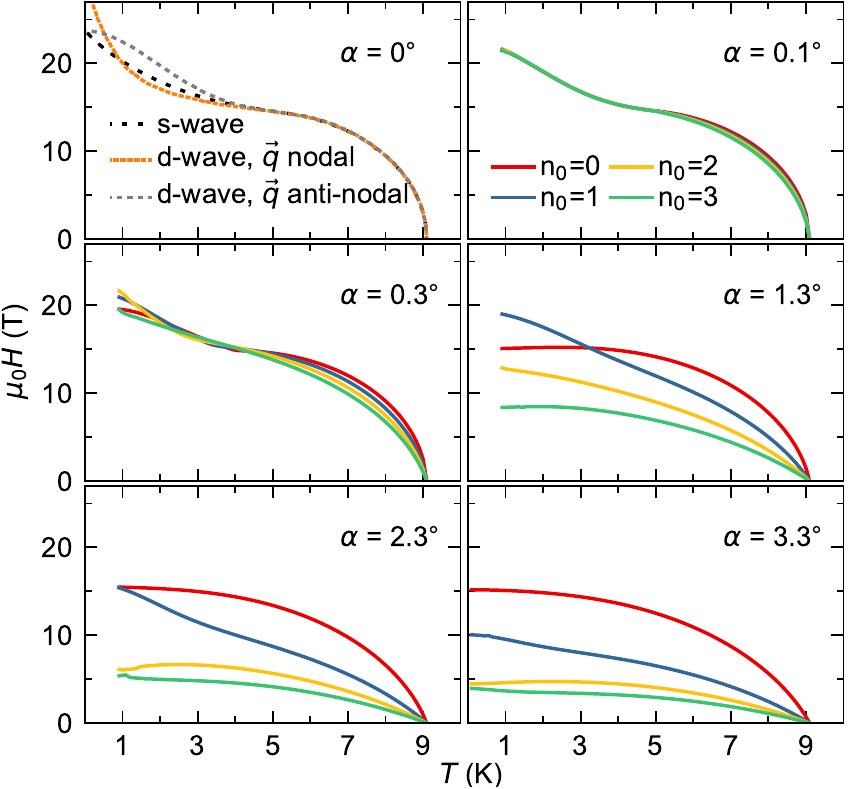}
	\caption{\label{Fig_S5} Calculated $H$-$T$ phase diagrams of $\kappa$-(ET)$_2$Cu(NCS)$_2$ and their variation with angle $\alpha$. Anisotropic $d$-wave pairing is assumed. For $\alpha=0$, the phase boundary resulting from a $d_x^2-y^2$-order parameter symmetry is shown for an FFLO modulation vector $\vec{q}$ pointing in nodal as well as anti-nodal direction and is compared to the FFLO solution assuming $s$-wave symmetry. For tilted field ($\alpha>0$), the four independent solution of Eq. (\ref{Eigenwert}) are plotted. The ground state at a temperature $T$ is given by the Landau level with the highest critical field.}
\end{figure}

Using Eqs. (\ref{Eigenwert}) to (\ref{off_d}), we calculated the critical fields of $\kappa$-(ET)$_2$Cu(NCS)$_2$ as function of the temperature for selected tilt angles $\alpha$ (Fig. \ref{Fig_S5}). Compared to Fig. \ref{Fig_S4}, the angular stability range of the high-field superconducting phase is slightly reduced for $d$-wave symmetry. However, qualitatively, one obtains identical behavior for $s$ and $d$-wave order parameters, proving that the claims of our paper do not depend on the underlying symmetry of the order parameter.

\section{Further effects that might influence the high-field phase diagram in $\kappa$-(ET)$_2$C$\mathrm{u}$(NCS)$_2$}
\subsubsection*{Spin-orbit coupling}
Concerning spin-orbit interactions, three effects might influence $H_{c2}$: (i) Spin-orbit scattering, which is negligible, since the organic superconductors are very clean systems. (ii) Periodic spin-orbit interactions with lattice ions, which is unlikely, since it scales with the proton number to the power of four and $\kappa$-(ET)$_2$Cu(NCS)$_2$ contains mostly light elements. (iii) Rashba-type interaction, which can occurs in layered systems from an asymmetry between the top and bottom face, as it was, for instance, treated in Ref. \cite{Zwicknagl2017}. However, this interaction would cause an anisotropy of $H_{c2}$ with respect to the magnetic-field direction within the plane. Although the latter effect cannot be completely ruled out, it seems unlikely, since a pronounced in-plane anisotropy has never been reported for $\kappa$-(ET)$_2$Cu(NCS)$_2$ \cite{Lortz2007,Agosta2012,Bergk2011}.
\subsubsection*{Commensurability effects}
Commensurability effects could, in principle, exist between the FFLO modulation and the Josephson vortices \cite{Bulaevskii2003,Uji2018}. The idea is that nodal lines of the FFLO modulation provide a pinning potential for the coreless Josephson vortices. While this effect will have a strong influence on the electrical resistance of the sample, it cannot account for the strong signatures in specific heat. 
\subsubsection*{Strong-coupling effects}
As discussed in our manuscript, $\kappa$-(ET)$_2$Cu(NCS)$_2$ is a strong-coupling superconductor. Typically, the strong-coupling leads to an enhancement of the critical fields compared to the BCS expectations \cite{Ashauer1987}. This effect is included in our discussion.
\subsubsection*{Interlayer coupling}
Strong interlayer coupling causes orbital effects to set in already for fields align in plane. However, as discussed in Ref. \cite{Lebed2018}, this is not the case for $\kappa$-(ET)$_2$Cu(NCS)$_2$.
\subsubsection*{Fluctuations}
The effect of fluctuations on the Abrikosov-like high-$n$ Landau states has not been studied so far. However, it has been shown that phase fluctuations affect a pure 2D FFLO system more seriously if it has a sinusoidally modulated order parameter than if it exhibits a more complex 2D modulation \cite{Shimahara1998}. Fluctuations might, therefore, stabilize higher-order Landau states further. 

\section{Specific-heat measurements}
\begin{figure*}[tb]
	\includegraphics{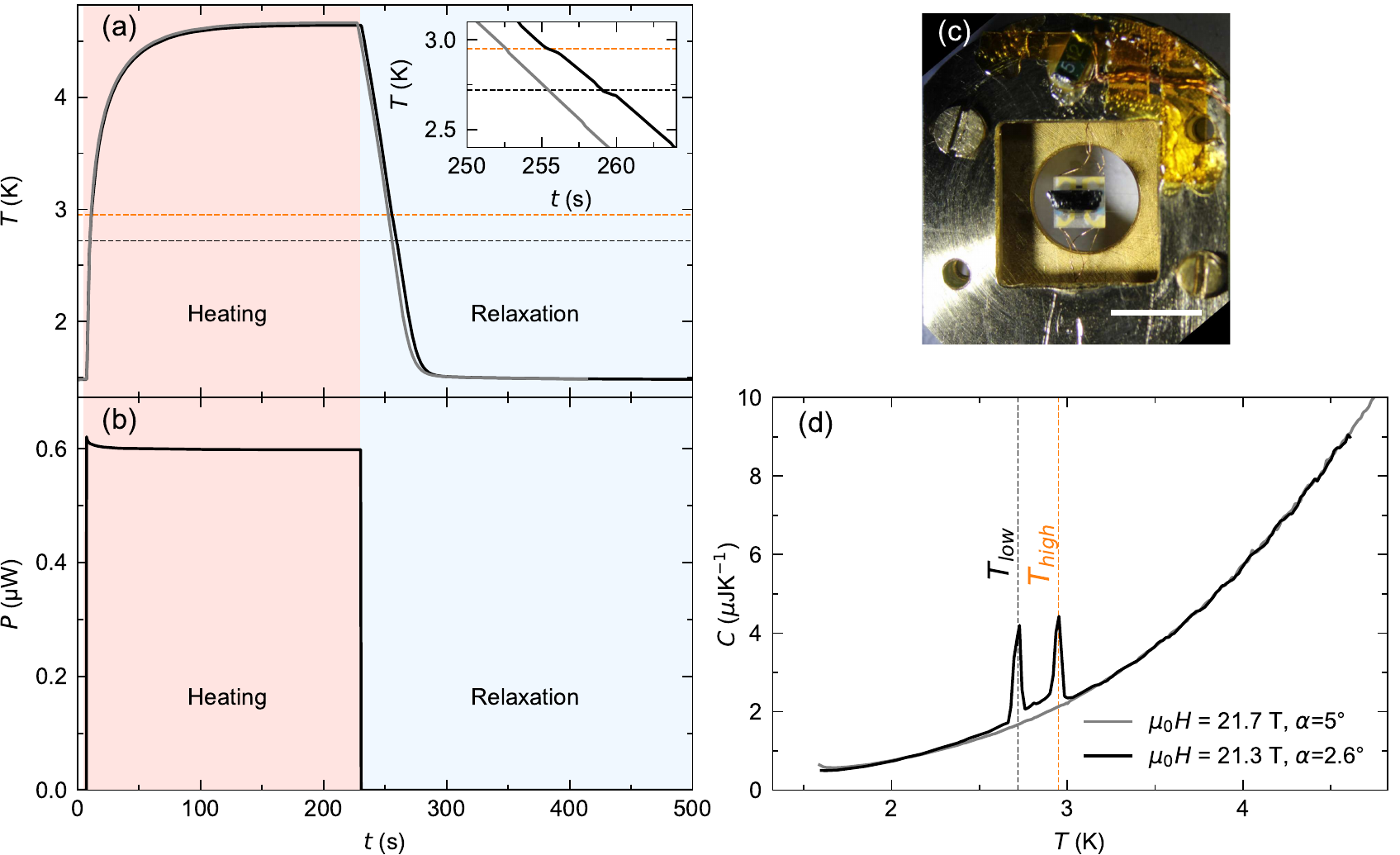}
	\caption{\label{Fig_S1}The continuous heat capacity method, shown exemplarily for data measured at \SI{21.3}{\tesla} and $\alpha=\SI{2.6}{\degree}$ [black lines in (a) and (d)]. Data for \SI{21.7}{\tesla} and $\alpha = \SI{5}{\degree}$ (gray lines), where the sample is in the normal-conducting state for all temperatures, are shown for comparison. (a) Temperature change during heating and relaxation. Inset: Enlarged view close to the transitions at $T_{low}$ and $T_{high}$. (b) Heater power during these measurements. (c) Photo of the modified heat-capacity puck. The scale bar corresponds to \SI{5}{\milli\meter}. (d) Heat capacity calculated from the data presented in (a) and (b) using Eq. (\ref{Eq1}).}
\end{figure*}

We attached the sample to a modified  $^3$He heat-capacity puck (Quantum Design) with a small amount of \mbox{Apiezon N} grease. In order to stabilize the sapphire platform of the puck, while further decreasing its thermal conductivity $\kappa_{th}$ to the bath, the leads connecting the platform were replaced by eight \SI{30}{\micro\meter} Manganin wires and the platform was additionally fixed to the center of the puck using \SI{10}{\micro\meter} Nylon threads [compare Fig. \ref{Fig_S1}(c)]. The puck was screwed to an Attocube ANRv51/LT piezo-driven rotator and placed into a \SI{22}{\tesla} Oxford Instruments cryomagnet with a \mbox{HelioxVL} $^3$He insert. The rotator allows for \SI{360}{\degree} rotation perpendicular to the magnetic field with an experimentally obtained reproducibility of a few \SI{0.01}{\degree}. The rotation axis was approximately parallel to the crystallographic $b$ axis. 

The heat capacity $C$ of the sample was measured using a continuous relaxation method which enables fast measurements, generates a high point density in $C$, allows to determine hystereses, and is particularly sensitive to first-order transitions \cite{Wang2001,Beyer2012}. Figure \ref{Fig_S1} illustrates the measurement procedure, in particular, the change of temperature [Fig. \ref{Fig_S1}(a)] and heater power [Fig. \ref{Fig_S1}(b)] during a data-acquisition cycle. Starting from the bath temperature $T = T_\mathrm{0}$, the platform with the sample on top is heated to $T\approx3T_\mathrm{0}$. Once a steady temperature is reached, the heater power $P$ is switched off, starting the relaxation of $T$ back to $T_\mathrm{0}$. For the heating ($T_0\rightarrow 3T_0$) as well as for the relaxation ($3T_0\rightarrow T_0$), the specific-heat $C$ of the sample in the range $T_\mathrm{0}< T< 3T_\mathrm{0}$ can be calculated from the time derivative of $T$, the heater power, the predetermined addenda, and the thermal conductance $\kappa_\mathrm{th}$ \cite{Wang2001}. For a discrete set of times $t_i$ and corresponding temperatures $T_i$, one obtains $C$ via
\begin{equation}
\begin{split}\label{Eq1}
C(\bar{T}_i) &= \frac{P\left(t_{i}\right)-\bar{\kappa}\cdot\left(\bar{T}_i-T_0\right)}{T_{i+1}-T_i}\left(t_{i+1}-t_{i}\right)\mathrm{ ,} \\
\bar{\kappa} &=\frac{1}{\bar{T}_i-T_0}\cdot\int_{T_0}^{\bar{T}_i}\kappa_{th}\left(T'\right)\mathrm{d}T',
\end{split}
\end{equation}
in which $\bar{T}_i=(T_i+T_{i+1})/2$. Figure \ref{Fig_S1}(d) presents the heat capacity calculated for the two measurements shown in Fig. \ref{Fig_S1}(a). For the \SI{21.3}{\tesla} data, the transitions arising at $T_{low}$ (black dashed line) and $T_{high}$ (orange dashed line) are already seen as kinks in the relaxation curves [inset Fig. \ref{Fig_S1}(a)]. Unless noted differently, each $C$ curve presented in our paper is the mean of five measurements.
\newpage
\section{Data measured at \SI{19.5}{\tesla} and \SI{21.7}{\tesla}}
\begin{figure*}[tb]
	\includegraphics{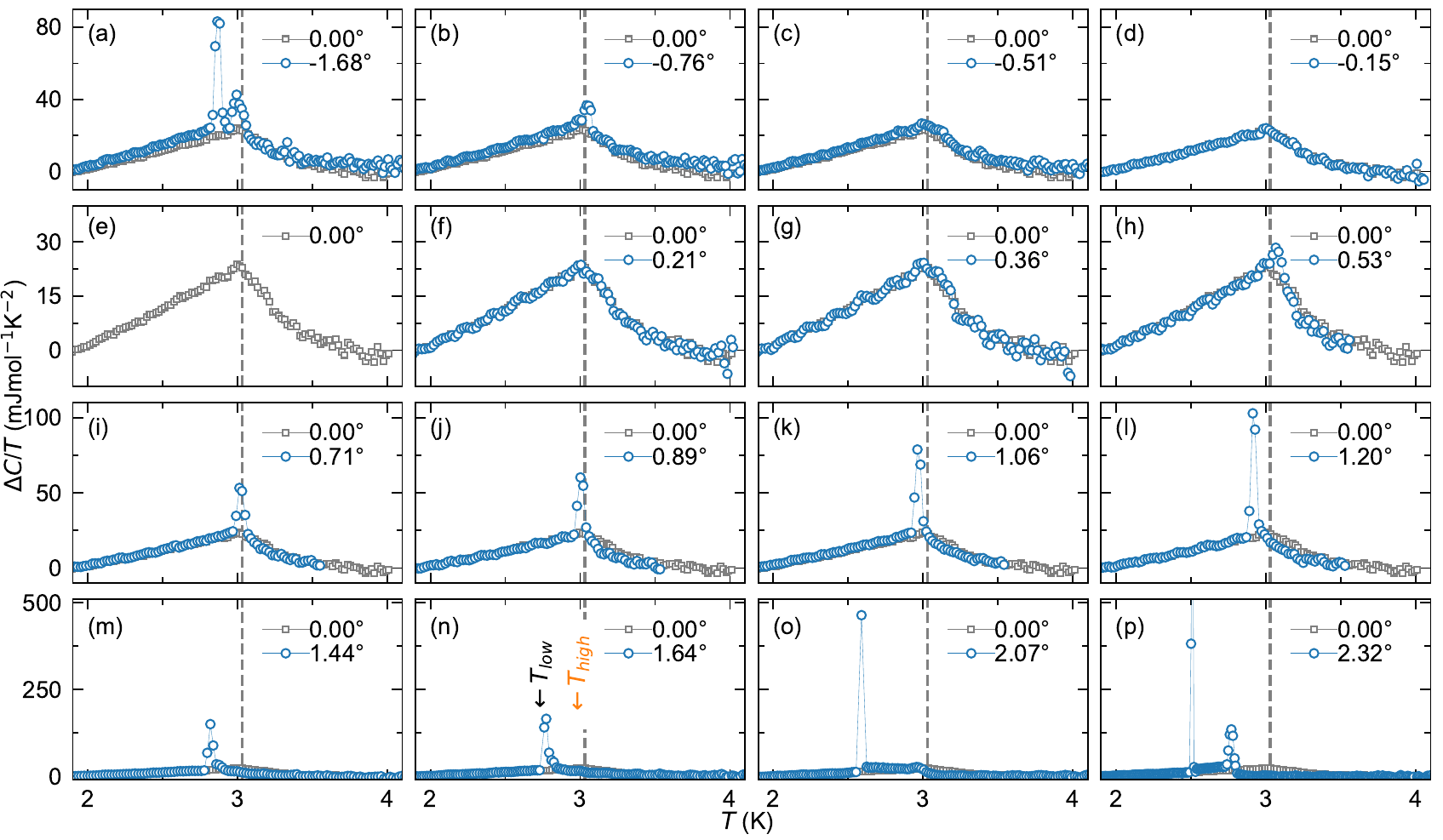}
	\caption{\label{Fig_S2} Specific heat of $\kappa$-(ET)$_2$Cu(NCS)$_2$ at \SI{21.7}{\tesla} for selected angles $\alpha$. The gray dashed line marks $T_{c}$ at \SI{21.7}{\tesla} and $\alpha=\SI{0}{\degree}$. Note the different scalings in each row.}
\end{figure*}
\begin{figure}[tb]
	\includegraphics{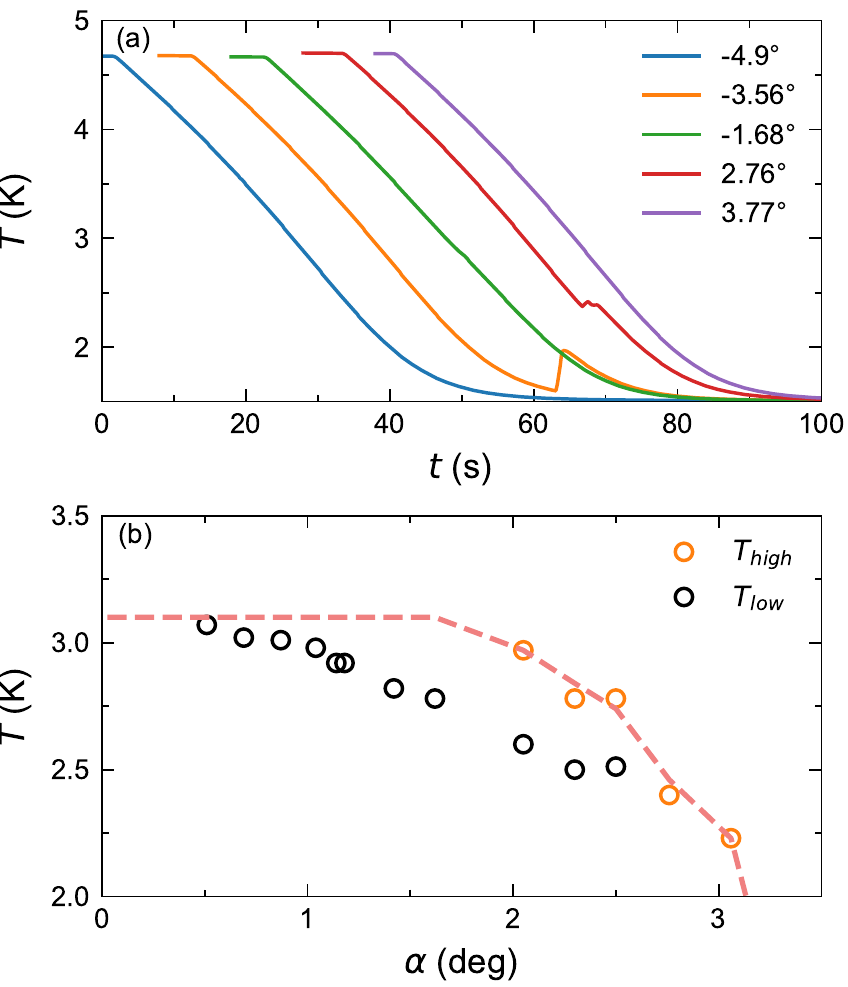}
	\caption{\label{Fig_S3} (a) Thermal-relaxation curves for selected $\alpha$ at \SI{21.7}{\tesla}. Clearly, pronounced supercooling effects are seen for negative as well as positive angles. (b)  $T$-$\alpha$ phase diagram of $\kappa$-(ET)$_2$Cu(NCS)$_2$ at \SI{21.7}{\tesla}.}
\end{figure}
\begin{figure*}[b]
	\includegraphics{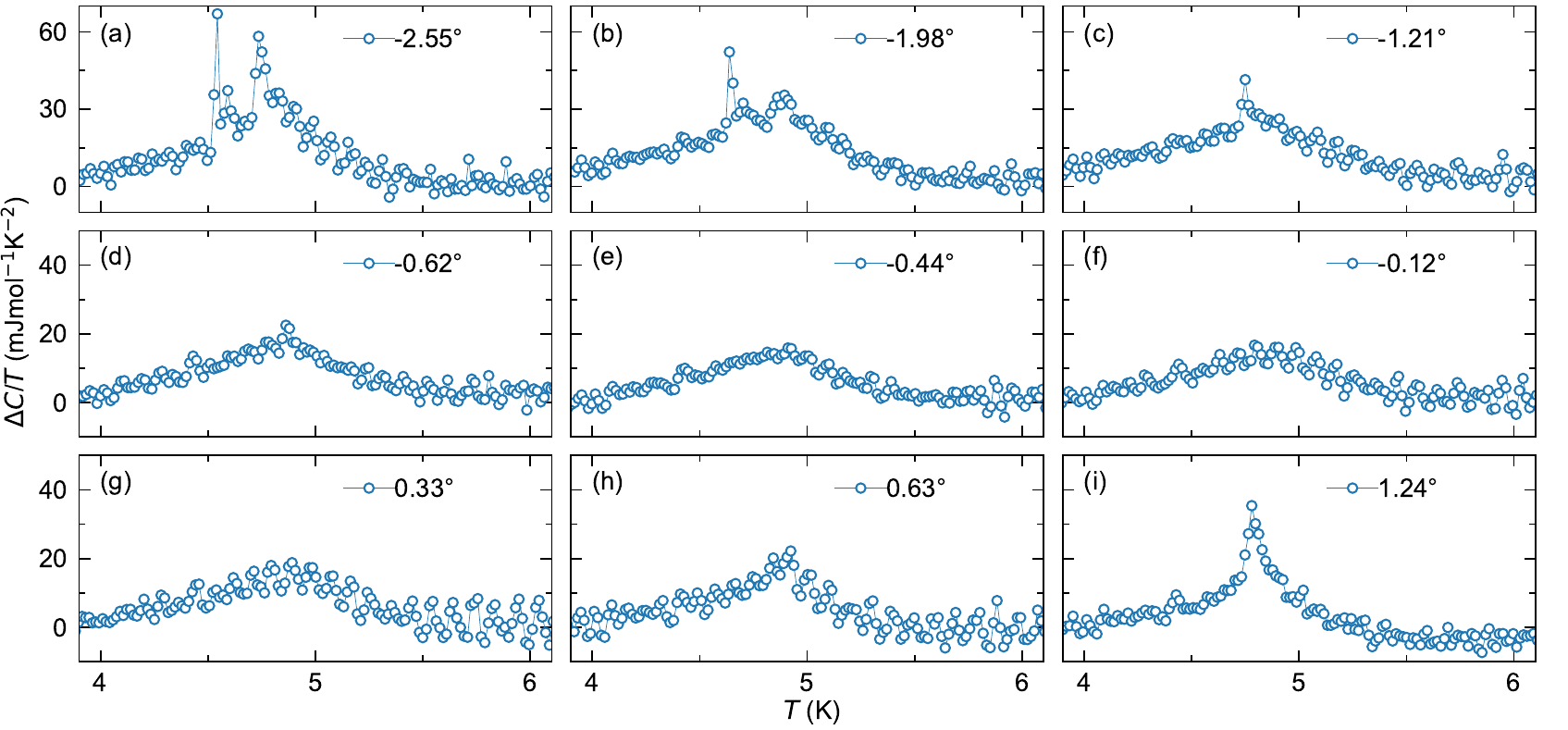}
	\caption{\label{Fig_S2B} Specific heat of $\kappa$-(ET)$_2$Cu(NCS)$_2$ at \SI{19.5}{\tesla} for selected angles $\alpha$. Note the different scalings in the first row. The data quality is reduced compared to Fig. \ref{Fig_S2}, due to a lower number of performed scans.}
\end{figure*}

In order to validate if the observed superconducting transitions follow the expected suppression with field, we repeated the angular-resolved specific-heat experiments, discussed in the main part for $\mu_0H=\SI{21.3}{\tesla}$, for slightly higher and lower fields. Figures \ref{Fig_S2}(a)-(p) show the specific heat of $\kappa$-(ET)$_2$Cu(NCS)$_2$ at \SI{21.7}{\tesla} for selected positive as well as negative angles $\alpha$. Additionally, relaxation curves are plotted in Fig. \ref{Fig_S3}(a) for selected angles. The curves are similar to the data obtained at \SI{21.3}{\tesla} (Fig. 3 in the main text), with clear non-equilibrium supercooling effects. We would like to emphasize two observations which prove the intrinsic nature of the measured first-order transition in $\kappa$-(ET)$_2$Cu(NCS)$_2$: (i) The features at $T_{high}$ and $T_{low}$ appear symmetrically around $\alpha=\SI{0}{\degree}$ verifying that the transitions are only related to the orientation of the applied field to the ET planes. (ii) As expected, $T_c$, $T_{high}$, and $T_{low}$ are slightly reduced compared to \SI{21.3}{\tesla}. However, the obtained phase diagram [Fig. \ref{Fig_S3}(b)] is qualitatively identical to the one presented in Fig. 4 in the main text. Comparable behavior is found for \SI{19.5}{\tesla} (Fig. \ref{Fig_S2B}), for which $T_{low}$ and $T_{high}$ are slightly higher. 
\newpage

\bibliography{20220313_kappa_ET_bibtex}